# Quantum complexity classes

M.Sc. thesis
by

Tereza Tušarová

Supervisor: Prof. Dr. H. M. Buhrman (CWI)
Second reader: Dr. F. van Raamsdonk (VU)

This material is the final thesis of the one year International Master's Program, supported by the Vrije Universiteit, Amsterdam. I wish to thank my supervisor Harry Buhrman for fruitful discussions. I thank my second reader Femke van Ramsdonk for reading the draft of this thesis thoroughly.

# Contents

**4** CONTENTS

# Introduction

Quantum computing is concerned with examining how the laws of quantum mechanics can be used to construct computation devices that would be possibly more powerful than the "classical ones", represented by Turing machines. As was pointed out in [9], computational devices behaving according to the quantum mechanics are not captured by the classical Church-Turing thesis stating that any practically realizable computational device can be simulated by a Turing machine. The reason is that the model of Turing machines is built implicitly assuming the laws of classical (e.g. nonquantum) physics hold, through it may seem that in the model there is no physics involved.

Thus, if we believe that the quantum theory is correct, we have to generalize the thesis to: "Any practically realizable computational device can be simulated by a quantum mechanical device." This automatically contains the classical Church-Turing thesis, since huge enough quantum systems may be described by the laws of classical physics with practically infinite precision. This generalized thesis gave rise to interest in quantum computing.

This research field entered much into fashion mainly thanks to Peter W. Shor, who published in 1994 his algorithm that can *efficiently* factor a given number [14]. Here the word efficiently means in polynomial time with respect to the length of the input. It's exponentially faster than any other known algorithm for a deterministic Turing machine. The problem of finding a factor of a given number is the core of many cryptographic systems, where it is assumed that to factor a large number is practically impossible in a reasonable time. It is believed to be a one-way problem: Multiplying up few numbers is very easy and can be done efficiently, while any known algorithm for a Turing machine requires exponential time.

After the algorithm, nowadays very famous, appeared, there has been a big rush for other algorithms showing an exponential speedup over classical algorithms. Surprisingly, until now, no other quantum algorithm [1] in the nonrelativized (eg. without oracles) setting that would offer such a speedup is known. Furthermore, there is not any physical realization known, that would allow to build computers with an unbounded amount of quantum bits without an exponential slowdown. In any case, before any serious attempt to build a quantum computer the question "How useful could a quantum computer be?" must be answered.

The class $BQP$ that contains all problems efficiently computable on a quantum computer should be well understood and placed into to the right place between other well-known traditional complexity classes like $NP$, $BPP$, etc. However nowadays, even fundamental questions like "Does $BQP$ contain NP-complete problems?" and "Does $BQP$ belong to the polynomial hierarchy?" still remain unanswered. Nor the lower bounds nor the upper bounds of the complexity of the class $BQP$ are very tight. A possible reason can be that finding answers to such fundamental questions about the class $BQP$ would immediately answer other fundamental questions from classical complexity theory that are considered to be very hard. For example, showing $BPP \neq BQP$ would immediately imply that $P \neq PSPACE$. Another reason might be that, through exactly formalized it is, the class $BQP$ is somehow counterintuitive and hard to work with. Intuitively, it would be easier to find relations between two quantum classes than between one conventional and one quantum class. A possible approach to go round these problems and still provide a better insight to the complexity of the class $BQP$ is to define a new bunch of classes somehow inspired by the $BQP$ class, place them to the conventional class hierarchy and then show some interesting properties of them together with their relation to $BQP$. One of such bunches is suggested here.

---

[1] or more precisely an algorithm based on a different idea which is here the quantum Fourier transform

The aim of this paper is to first review the model of quantum computation, both from the view of Quantum Turing machines and the view of Quantum circuits. The relation between those two is sketched and also the phenomenon of entanglement and its importance are discussed. The main work comes after that. We introduce new quantum complexity classes related to the $BQP$ class and show their properties.

This thesis is organized as follows: In chapter 2 we define the quantum Turing machine and show some of its properties. In the next chapter, we introduce quantum circuits. Chapter 4 shows the relations between quantum circuits and quantum Turing machines. We focus on comparing the power of the two models. The main results of this work show up in chapter 5. Therein, we define new complexity classes and show some interesting properties of them. The new hierarchy of classes is illustrated in section 5.7, Figure 5.7, together with some other well-known complexity classes.

In the text, we assume that the reader is familiar with basic linear algebra and with basic notion of complexity classes and Turing machines. We do not assume any kind of physics knowledge. Those, who are interested in the notion of Quantum Computing from a computer scientist point of view, can start from the beginning with chapter 2. Those, who are already familiar with Quantum Computing and are only interested in the new bunch of classes, can refer to chapter 5.

# Chapter 1

# Preliminaries

In this chapter we summarize definitions and theorems known in linear algebra that are used in this text.

## 1.1 Hilbert space

We will need these spaces to be able work with quantum gates and quantum states. A Hilbert space, through its name may suggest otherwise, is a natural notion of a space that has all the nice properties we imagine and many times take for obvious. We need to have an inner product defined in order to measure angles, to have a norm defined in order to choose only unit norm vectors and we need that space be complete, because we do not want to slip out of it when doing a limit. We formalize these properties bellow.

**Definition 1.1 (Inner product).** *An inner product over a vector space $V$ is a complex function of two arguments denoted by $\cdot$ and satisfying*

1. $\forall x \in V : x \cdot x \geq 0$

2. $x \cdot x = 0 \iff x = \vec{0}$

3. $\forall x, y, z \in V : (ax + by) \cdot z = a(x \cdot z) + b(y \cdot z)$

4. $\forall x, y \in V : x \cdot y = (y \cdot x)^*$

Each inner product induces a norm $|x|$ given by $|x| = \sqrt{x \cdot x}$

**Definition 1.2 (Inner product space).** *An inner product space is a vector space where an inner product $\cdot$ is defined.*

**Definition 1.3 (Perpendicular vectors).** *If for two vectors $x, y \in V$ for some inner product space $V$, it holds that $x \cdot y = 0$, we say that $x$ is perpendicular to $y$ and write $x \perp y$.*

**Definition 1.4 (Complete space).** *A space $H$ with a defined norm $|x|$ is complete under that norm iff for each sequence $\{x_i\}_{i=1}^{\infty}$ such that $x_i \in H$ and $\lim_{n,m \to \infty} |x_n - x_m| = 0$, it holds that $\lim_{n \to \infty} x_n = x \in H$.*

**Definition 1.5 (Hilbert space).** *A Hilbert space is an inner product space that is complete under the induced norm.*

**Definition 1.6 (Dual space).** *A dual space to an inner product vector space $V$ is a space $V^D$ spanned by linear functionals $x^D : V \to \mathbb{C}$ given by $x^D(y) = x \cdot y$. The space $V^D$ is also an inner product space, with an inner product defined as $(x^D \cdot y^D)(z) \stackrel{DEF}{=} (x \cdot y)$.*

**Dirac notation**  We write $|x\rangle$, the so called "ket-vector" to emphasize that $x$ is an element of a Hilbert Space. $\langle x|$, so called "bra", is element of dual Hilbert space. The inner product of $x$ and $y$ is symbolically written as $\langle x||y\rangle$, which is usually shortened to $\langle x|y\rangle$. If we take a linear operator $O$ and apply it onto a ket $|x\rangle$, we get a new vector $|Ox\rangle$ which we denote as $O|x\rangle$. We can then write the inner product of $y$ and $Ox$ as $\langle y|Ox\rangle \equiv \langle y|O|x\rangle$. If both $x$ and $y$ are base vectors, then the expression $\langle y|O|x\rangle$ is called a "matrix element" of the operator $O$, and also denoted as $O_{x,y}$. Having specified all its matrix elements, we have specified the linear operator itself. Thus, in the text, we will use the terms "operator" and "matrix" interchangeably, using the fact that we talk only about linear operators, since only these occur in quantum mechanics.

**Definition 1.7 (Superposition).** *Let us have a Hilbert space $H$ and its base $\{_{i=1}^{n}|e_i\rangle\}$. Then each vector $v$ from $H$ can be written as $v = \sum_{i=1}^{n} a_i|e_i\rangle$. If more than one coefficient $a_i$ is nonzero, we say that $v$ is a* superposition *of corresponding $e_i$'s. Often, we use the word "superposition" to indicate that $v$ is not a base vector.*

## 1.2  Tensor product

**Definition 1.8 (Tensor product of two Hilbert spaces).** *A tensor product of two Hilbert spaces $U$ and $V$ is a vector space $W$, denoted as $W = U \otimes V$, spanned by all possible pairs of vectors*

$$\{u \otimes v | u \in U, \ v \in V\}$$

*Its base $W_b$ is*

$$W_b = \{u \otimes v | u \in U_b, \ v \in V_b\}$$

*where $U_b, V_b$ are bases of $U$ and $V$ respectively. The space $W$ is again a Hilbert space, with an inner product $\cdot$ defined as*

$$a \otimes b \cdot c \otimes d = (a \cdot c)(b \cdot d)$$

*This also automatically defines a tensor product of two vectors and a tensor product of two linear operators.*

Many times in this text, we will omit the $\times$ sign and write $|ab\rangle$ instead of $|a\rangle \otimes |b\rangle$ where no confusion may occur. We illustrate the notion in the following example:

**Example 1.1 (Tensor product).** *Let us have two Hilbert spaces $U$ and $V$ with bases*

$$U_b = \{|0\rangle, |1\rangle\}$$
$$V_b = \{|0'\rangle, |1'\rangle\}$$

*Then the space $W = U \otimes V$ is spanned by base*

$$W_b = \{|00'\rangle, |01'\rangle, |10'\rangle, |11'\rangle\}$$

*If we have vectors $u \in U$, $v \in V$ defined as*

$$|u\rangle = a|0\rangle + b|1\rangle$$
$$|v\rangle = c|0'\rangle + d|1'\rangle$$

*the vector $w = u \otimes v \in U \otimes V$ reads*

$$w = ac|00'\rangle + ad|01'\rangle + bc|10'\rangle + bd|11'\rangle$$

If we have two linear operators $A$ and $B$ defined on spaces $U$ and $V$ respectively, with their matrix representations $A_{ij}$ and $B_{ij}$, their product $C = A \otimes B$ working on space $U \otimes V$ has matrix $B_{(ij)(kl)} = A_{ik}B_{jl}$. For example, if

$$A = \frac{1}{\sqrt{2}} \begin{pmatrix} 1 & 1 \\ 1 & -1 \end{pmatrix}$$

and

$$B = \begin{pmatrix} 0 & -1 \\ 1 & 0 \end{pmatrix}$$

then

$$C = A \otimes B = \begin{pmatrix} A_{11}B_{11} & A_{11}B_{12} & A_{12}B_{11} & A_{12}B_{12} \\ A_{11}B_{21} & A_{11}B_{22} & A_{12}B_{21} & A_{12}B_{22} \\ A_{21}B_{11} & A_{21}B_{12} & A_{22}B_{11} & A_{22}B_{12} \\ A_{21}B_{21} & A_{21}B_{22} & A_{22}B_{21} & A_{22}B_{22} \end{pmatrix} = \frac{1}{\sqrt{2}} \begin{pmatrix} 0 & -1 & 0 & -1 \\ 1 & 0 & 1 & 0 \\ 0 & -1 & 0 & 1 \\ 1 & 0 & -1 & 0 \end{pmatrix}$$

## 1.3 Qubits and gates

Quantum mechanics tells us that a state of any quantum system is always described by a unit norm ket vector. Thus, all vectors we will work with will have a unit norm:

$$\forall x : \langle x|x \rangle = 1$$

**Definition 1.9 (Qubit).** *A qubit is a unit norm element of a Hilbert space of dimension 2. If we label the base vectors $|0\rangle$ and $|1\rangle$, a qubit is of the form $\alpha|0\rangle + \beta|1\rangle$ where $|\alpha|^2 + |\beta|^2 = 1$.*

**Definition 1.10 (Hermitian conjugate).** *A hermitian conjugate of an operator $O$ is such an operator $O^\dagger$ for which it holds:*

$$\forall x, y : \langle x|Oy \rangle = \langle O^\dagger x|y \rangle$$

**Definition 1.11 (Inverse operator).** *An operator $O^{-1}$ is inverse operator of operator $O$ iff*

$$\forall x : |x\rangle = |O^{-1}Ox\rangle = |OO^{-1}x\rangle$$

**Definition 1.12 ($\delta$ matrix).** *The $\delta$ matrix is defined as*

$$\delta_{ij} = \begin{cases} 1 & \text{if } i = j \\ 0 & \text{otherwise} \end{cases}$$

*Many times, we will write the $\delta$ matrix simply as 1.*

**Definition 1.13 (Separable matrix).** *A matrix $M$ is separable if it can be written as $M = A \otimes B$ for $A$ and $B$ of dimension at least 2.*

Quantum mechanics also tells us, that the time evolution of a quantum system is unitary. Thus, in this text, we will encounter only unitary operators.

**Definition 1.14 (Unitary operator).** *An operator $O$ is unitary iff $O^\dagger = O^{-1}$.*

**Observation 1.1.** *Unitary operators preserve the inner product.*

*Proof.* We have
$$\forall x, y : \langle Ux|Uy\rangle = \langle U^\dagger Ux|y\rangle = \langle U^{-1}Ux|y\rangle = \langle x|y\rangle$$

$\square$

**Lemma 1.2 (Conditions for unitarity).** *A matrix $U$ is unitary iff all its columns and rows have a unit norm and are mutually orthogonal. Formally:*

$$\sum_k \langle k|U|y_1\rangle^* \langle k|U|y_2\rangle = \delta_{y_1, y_2}$$

*Proof.* We need to show that: $UU^\dagger = U^\dagger U = 1$. We have

$$\delta_{y_1, y_2} = \sum_k \langle k|U|y_1\rangle^* \langle k|U|y_2\rangle = \sum_k \langle y_1|U^\dagger|k\rangle \langle k|U|y_2\rangle = \langle y_1|U^\dagger U|y_2\rangle$$

and thus $U^\dagger U = 1$. This leads to

$$UU^\dagger = (UU^\dagger)UU^{-1} = U(U^\dagger U)U^{-1} = UU^{-1} = 1$$

$\square$

# Chapter 2

# Quantum Turing machine

In this chapter, we will treat the quantum Turing machine and compare its properties with those of classical Turing machines. First, we will review the classical Turing machine and its basics properties. Then, we will jump to its generalized version: the PTM. At last, we will define a quantum Turing machine in an analogous way and point out at the differences to the other types of Turing machines.

## 2.1 Classical Turing machines

The Turing machine serves as a model of a general computational device. As the Church-Turing thesis says, anything a realizable physical device can compute, the Turing machine can compute. The basic one, classical deterministic Turing machine is the simplest of all Turing machines. It behaves according to well defined strict rules, which is also what our conventional computers do: They process an exact program stored in a memory, step by step. On the other side, quantum computers, if ever built, will be able to do much more things. This obviously means that also the model of such computers, the quantum Turing machine will be somehow different. To show the logic of its definition, we will first start with the classical Turing machine and modify it step by step to finally arrive at the definition of the quantum Turing machine at the end of this chapter.

### 2.1.1 Classical deterministic Turing machine

A Turing machine is a device that moves along a possibly infinite tape. Is is equipped with a head that can read symbols from the tape and write symbols to the tape. The set of possible symbols is finite. The tape is divided into cells and the head can move to the neighbor cells in both directions (left or right) in a single step. The head also has a finite set of states. The current state of the head and the symbol read from the current position determine the state the head will turn to, the symbol it will write to the tape and the direction it will move to. The behavior is encoded in a transition function $\delta$:

$$\delta(\sigma, q) = (\sigma', q', D)$$

where

$\sigma$ is the symbol read from the current position on the tape

$q$ is the state the head is in

$\sigma'$ is the symbol that will be written to the current position on the tape

$q'$ is the state the head will turn to

$D$ is the direction the head will move to, its either $D = L$ standing for "left" or $D = R$ standing for "right"

Since we have only a finite set of possible symbols and a finite set of possible states, the definition of the $\delta$ function is also finite. The only infinite thing is the tape. That corresponds to our experience: We want to have a device that can handle inputs of possibly unbounded size without a need to alter the device in any way.

Initially, the input is written on the tape and the head is in its special initial state. Once the task is done, the machine stops. This happens when the $\delta$ function is not defined on the corresponding symbol and state. Formally, the Turing machine is defined as follows:

**Definition 2.1 (Turing machine).** *A Turing machine (TM) is an ordered seven-tuple $(\Sigma, \Lambda, Q, q_i, A, F, \delta)$, where*

*$\Sigma$ is a finite set, called "alphabet", of all possible tape symbols.*

*$\lambda \in \Sigma$ is an identified symbol called a "blank symbol"*

*$Q$ is a finite set, called "set of states"*

*$q_i \in Q$ is the initial state*

*$A \subseteq Q$ is a set of accepting states*

*$F \subseteq Q$ is a set of final states*

*$\delta : \Sigma \times Q \to \Sigma \times Q \times \{L, R\}$ is a transition function.*

*It operates on a tape of cells indexed by $\mathbb{Z}$. A Turing machine internally stores:*

- *$q_{current} \in Q$, initialized to $q_i$*

- *integer $i$, initialized to 0. $i$ is the position on the tape which will be read in the next step.*

*A computation consists of periodical repeating of steps that are determined by the $\delta$ function. A Turing machine is said to halt when it reaches a state from $F$. If it halts in some state in $F \cap A$, it is said to accept.*

Usually, a Turing machine has various tapes, not only one. However, its known that each multi-tape Turing machine can be simulated by a single-tape Turing machine with no more than a polynomial slowdown. Therefore, we restrict ourselves to single-tape Turing machines.

A Turing machine can be treated either as an *acceptor* or as a *transducer*. An acceptor is a machine that does not write any output onto the tape, the only information it tells us is the state it halted in. A transducer, in a contrary, writes the output onto the tape, while the state it halts in may be arbitrary. Formally:

**Definition 2.2 (Transducer).** *A transducer is a Turing machine that uses its tape as an output tape. When the machine halts, the output is written on the tape, beginning from the cell indexed by zero to the first blank symbol on the right. A transducer thus computes a function $f : (\Sigma - \lambda)^* \to (\Sigma - \lambda)^*$. If the transducer does not halt on some inputs, the function is not defined there.*

**Definition 2.3 (Acceptor).** *An acceptor is a Turing machine that does not write any output onto the tape. When it halts, it either accepts or not, depending on whether the final state is in the set $A$.*

**Definition 2.4 (Accepting a language).** *We say that a Turing machine $M$ accepts a language $L$ if it always halts and for each input $x$:*

*if $x \in L$ then the machine $M$ accepts on input $x$,*

*if $x \notin L$ then the machine $M$ does not accept on input $x$.*

*We then write $L = L(M)$.*

We have said that the number of states and the number of possible symbols and so also the transition functions are fixed and do not change with the length of the input. However, with the increasing lengths of the input, the number of computational steps may increase and so the number of cells the machine visits. To capture this, time and space complexity of a machine is defined.

**Definition 2.5 (Time complexity).** *Let $T(n)$ be a function $T : \mathbb{N} \to \mathbb{N}$ and let $M$ be a Turing machine that on each input of length $n$ proceeds at most $T(n)$ steps before halting, for some function $T(n)$. We then say that $M$ has time complexity $T(n)$.*

**Definition 2.6 (Space complexity).** *Let $S(n)$ be a function $S : \mathbb{N} \to \mathbb{N}$ and let $M$ $M$ be a Turing machine which on each input of length $n$ uses (writes to and reads from) at most $S(n)$ cells before halting, for some function $S(n)$. We then say that $M$ has space complexity $S(n)$.*

**Definition 2.7 (Configuration of a Turing machine).** *A configuration of a Turing machine is an ordered set of:*

- *the contents of the tape*
- *the current state*
- *the position of the head*

*The set of all configurations of a Turing machine $M$ will be denoted by $C(M)$.*

It is sometimes useful to think of a Turing machine as of a *transition matrix* $T$, transforming configurations into each other. If a configuration $c_1$ leads to another configuration $c_2$ in the next step, there is 1 on the position $T_{c_2, c_1}$. [1] Otherwise there is zero. Because the tapes are infinite, the matrix is infinite-dimensional. However, if we know that the time complexity of a Turing machine is $T(n)$, we may for fixed $n$ have a finite dimensional matrix cutting the tapes at the distance $T(n)$ from the initial position on both sides. The size of the matrix is then $|Q| \cdot (2T(n) + 1) \cdot |\Sigma|^{2T(n)+1}$ which is the number of configurations on a tape of length $2T(n) + 1$.

**Definition 2.8 (Transition matrix).** *Let us have a deterministic Turing machine with a time complexity $T(n)$. A transition matrix for the length $n$ is a square matrix $T$ of size $|Q| \cdot (2T(n) + 1) \cdot |\Sigma|^{2T(n)+1} \times |Q| \cdot (2T(n) + 1) \cdot |\Sigma|^{2T(n)+1}$ where*

$$T_{c_i, c_j} \stackrel{DEF}{=} \begin{cases} 1 & \text{if the machine goes from } c_j \text{ to } c_i \text{ in one step} \\ 0 & \text{otherwise} \end{cases}$$

It should be pointed out that for most matrices containing only zeros and ones, no corresponding Turing machine exists. The reason is that each Turing machine has a finite description of bounded size. A constant size description can only generate constant amount of matrices. Using this definition, each deterministic Turing machine can be expressed as a family of transition matrices $T_1, T_2, \ldots$ where $T_i$ is a transition matrix for length $i$.

### 2.1.2 Deterministic Turing machines with oracles

Oracles are abstract devices answering yes-no questions in a unit time. Its natural to ask then, if we supply a Turing machine with this kind of information, what more can it compute? We will first formalize the situation of a Turing machine with oracle.

---
[1] Note that the indices are in reverse order.

**Definition 2.9 (Oracle).** *An oracle over an alphabet $\Sigma$ is an arbitrary set of words from this alphabet, which we can ask whether any $x \in \Sigma^*$ is there and get the answer in a unit time.*

Naturally, one would expect that an oracle Turing machine can do two different things: It either undergoes a transition given by its $\delta$ function or asks the oracle and accordingly changes its state. Nevertheless, to grasp this, we would have to change the definition of the $\delta$ function. Because of this, a more common approach is to introduce a special query state in which the oracle writes to the tape the answer for the word it finds on the tape and then the machine turns to another special post-query state. Obviously, the two approaches are equivalent. We review the second approach kind oracle Turing machine for completeness.

**Definition 2.10 (Oracle Turing machine).** *An oracle Turing machine is an ordered ten-tuple $(\Sigma, \Lambda, Q, q_i, A, \delta, q_q, q_a, y, n)$, where the first six items are the same as in the Definition 2.1 and*

$q_q \in Q$ *is a special query state*

$q_a \in Q$ *is a special post-query state*

$y \in \Sigma$ *is a special 'yes' symbol*

$n \in \Sigma$ *is a special 'no' symbol*

*It works with any oracle that has the same alphabet. When the machine enters the query state $q_q$, it asks the oracle for the word written on the tape beginning from the actual position to the first blank symbol on the right. Then the answer is written onto the actual position, the 'yes' symbol if the oracle answers "yes" and the 'no' symbol if the oracle answers "no". Then the machine enters into the post-query state $q_a$ and continues the computation as usual.*

### 2.1.3 Probabilistic Turing machine

**Definition 2.11 (Probabilistic Turing machine).** *A probabilistic Turing machine is a slightly modified Turing machine. Its transition function does not output only one triple of symbol written, new state and direction to move, but outputs a probabilistic distribution of such triples.[2] If a probabilistic Turing machine halts, each of possible outputs has a certain probability. We say that a probabilistic Turing machine produces a certain probability distribution. If a probability of halting in a state in $F \cap A$ is $p$, we say that the probabilistic Turing machine accepts with probability $p$.*

If we look at the finite version of transition matrix here, we see that it does not have only 1s and 0s, but may contain any number from the interval $[0, 1]$. It is a stochastic matrix where each column sums up to one. Instead of $\delta : \Sigma \times Q \to \Sigma \times Q \times \{L, R\}$, we will here talk about $\delta : Q \times \Sigma \times Q \times \Sigma \times \{L, R\} \to \mathbb{R}^+$. We define $\delta(q_2, \sigma_1, q_2, \sigma_2, D) \stackrel{\text{DEF}}{=} \alpha$ iff the probability of going from the state $q_1$, having read the symbol $\sigma_1$, to the state $q_2$, moving to the direction $D$ and writing the symbol $\sigma_2$ is $\alpha$.

**Definition 2.12 (Accepting a language on a probabilistic Turing machine).** *We say that probabilistic Turing machine $M$ accepts language $L$ with completeness $c$ and soundness $s$ iff over an input $x$*

*for $x \in L$, $M$ accepts with probability $p \geq c$,*

*for $x \notin L$, $M$ accepts with probability $p \leq s$.*

*Then we write $L = L(M, c, s)$*

---

[2]Strictly speaking, it chooses among different possibilities how to continue the computation, each having certain probability.

The following lemma tells us, that we may restrict ourselves to cases where there are only two computation paths possible, each with the probability one half. It can be found for example in [12].

**Lemma 2.1.** *For each probabilistic Turing machine $M$ with time complexity $T(n)$ and each $\epsilon > 0$, there exists a probabilistic Turing machine $M'$ with time complexity $T'(n)$ such that its possible probabilities are from the set $\{0, \frac{1}{2}, 1\}$ and for each language $L$ and numbers $c, s$ it holds $L(M, c, s) = L(M', c - \epsilon, s + \epsilon)$ and $T'(n) = p(T(n))$ for some polynomial $p(n)$.*

At each step of a probabilistic Turing machine where there are two possible computation paths, we may imagine that the machine behaves deterministically, but uses one random bit. At each computation of a time complexity $T(n)$, it may use at most $T(n)$ random bits. This leads to the following characterization of languages:

**Observation 2.2.** *For each language $L$, a probabilistic Turing machine $M$ with time complexity $T(n)$ and numbers $c, s$, such that $L = L(M, c, s)$, there exists a deterministic Turing machine $M_d$ such that*

*For $x \in L$, $P_{y:\text{length}(y)=T(n)}(M_d(x, y) accepts) \geq c$.*

*For $x \notin L$, $P_{y:\text{length}(y)=T(n)}(M_d(x, y) accepts) \leq s$.*

*where $P_{y:\text{length}(y)=T(n)}$ is the probability over all $y$'s whose length is $T(n)$. We may then write $L = L((M_d, \text{length}(T(n))), c, s)$.*

We put the definition of a nondeterministic Turing machine here for sake of completeness. Having already defined a probabilistic Turing machine, the nondeterministic Turing machine can be thought of being its special case.

**Definition 2.13 (Nondeterministic Turing machine).** *A nondeterministic Turing machine differs from a probabilistic Turing machine only in the condition of acceptation: We say that a nondeterministic Turing machine $M$ accepts a language $L$ iff*

*For $x \in L$, $M$ accepts with probability $p > 0$.*

*For $x \notin L$, $M$ accepts with probability $0$.*

*Then we write $L = L(M)$.*

## 2.2  Quantum Turing machine

A formal definition of a quantum Turing machine was first given in [7]. Here we will reformulate the definition to obtain a more convenient form. When defining a quantum Turing machine, we cannot straightforwardly apply the previous definitions of halting onto the quantum case. The quantum mechanical nature tells us the time evolution of a quantum system has to be unitary. In other words, if we express the state of a given system as a vector from a Hilbert space, each subsequent state of this system is obtained by applying a unitary operator to that vector. The operator is given by the concrete physical situation.

If we think of configurations as of a base of a Hilbert space, we get the condition that the transition matrix has to be unitary, because its in fact an operator that tells us in which state the system (here the machine and the tape) will be in the next step.

**Definition 2.14 (Quantum Turing machine informally).** *A quantum Turing machine (QTM) is similar to a probabilistic Turing machine, with the following differences: Here the coefficients are not probabilities, but complex numbers called* amplitudes. *At each step, the squares of norms of amplitudes of possibilities sum up to one. For each length of input, the corresponding transition matrix has to be unitary.*
*We say the machine halts if all branches enter the final state. The output is then written on the tape, beginning from the initial position to the first blank symbol. The probability of outputting a configuration is a squared norm of the corresponding amplitude.*

It should be emphasized that the transition matrix now contains complex numbers. Now, we may think of configurations as of orthonormal base vectors of some Hilbert space $H$. Then, at each step the $\delta$ function output corresponds to some vector $v$ from $H$, $v = \sum a_i c_i$, where $c_i$ are configurations (so base vectors), and $a_i$ are corresponding amplitudes. The definition of quantum Turing machine tells us that the sum of norms of $a_i$, $\sum a_i^* a_i$ has to be 1. We will thus work only with unit norm vectors. We will also take a profit of the Dirac notation and write

$$\delta(c) = \sum_{i=1}^{k} a_i |c_i\rangle$$

to indicate that from configuration $c$, there are $k$ possible resulting configurations $c_i$, each with respective amplitude $a_i$. Now, we can define quantum Turing machine more formally as:

**Definition 2.15 (Quantum Turing machine formally).** *A quantum Turing machine is an ordered six-tuple $(\Sigma, \Lambda, Q, q_i, q_f, \delta)$, where*

- *$\Sigma$ is a finite set, called "alphabet", of all possible tape symbols. We assume it equals $\{0, 1, \Lambda\}$.*

- *$\Lambda \in \Sigma$ is called a "blank symbol"*

- *$Q$ is a finite set, called "set of states"*

- *$q_i$ is the initial state*

- *$q_f$ is the final state*

- *$\delta : \Sigma \times Q \to H$ is a transition function and $H$ is a Hilbert space spanned by base vectors corresponding to triples from $\Sigma \times Q \times \{L, R\}$,*

*and the corresponding finite transition matrix is unitary for all lengths of input.*

The acceptation of a language on a quantum Turing machine has to be redefined, due to the change in the interpretation of probabilities:

**Definition 2.16 (Acceptation of a language).** *We say that quantum Turing machine $M$ accepts language $L$ iff there is a configuration $a(x)$ such that for each $x$ the machine $M$ halts in a configuration $\sum_i a_i |c_i\rangle$ such that*

- *If $x \in L$ then $\sum_i a_i |\langle a(x) | c_i \rangle|^2 > \frac{2}{3}$*

- *If $x \notin L$ then $\sum_i a_i |\langle a(x) | c_i \rangle|^2 \leq \frac{1}{3}$*

The function $a(x)$ in the definition above tells us, which configuration will indicate $x \in L$ for each $x$. Then we want that the probability of outputting such configuration is bigger than two thirds for $x \in L$ and smaller or equal to one third for $x \notin L$.

As for deterministic and probabilistic Turing machines, it was shown that we suffice with rational entries in the transition matrix, here thus amplitudes.

**Lemma 2.3 (Constant number of amplitudes).** *For each quantum Turing machine $M$ with time complexity $T(n)$ there exists a quantum Turing machine $M'$ that has only the numbers $0, \pm\frac{3}{5}, \pm\frac{4}{5}, 1$ as amplitudes, accepts the same language and its time complexity is $p(T(n))$ for some polynomial $p$.*

*Proof.* In [1]. □

### 2.2.1 Oracle quantum Turing machines

As in the classical case, we will want to provide quantum Turing machines with oracles. However, to incorporate an oracle into the notion of a quantum Turing machine, we need that the query process does not violate the unitarity condition (see Def. 2.15) of a quantum Turing machine. That means that the the computation has to be reversible - no information can disappear from the system. This for example implies that the answer cannot be written over a non-blank symbol that was not previously copied to somewhere else etc. We will thus assume that oracle changes the current bit if the answer is "yes" and leave in unchanged if the answer is "no". In other words, we will assume that if the contents of the tape to the first blank symbol on the right is $|x, b\rangle$, then in the post-query state it is $|x, b \oplus I(x)\rangle$ where $I(x) = 1$ if $x \in O$ and $I(x) = 0$ otherwise. For sake of completeness, we also write here the formal oracle quantum Turing machine definition.

**Definition 2.17 (Oracle quantum Turing machine formally).** *A Turing machine is an ordered eight-tuple $(\Sigma, \Lambda, Q, q_i, q_f, \delta, q_q, q_a)$, where*

$\Sigma = \{0, 1, \Lambda\}$ *is a finite set, called "alphabet".*

$\Lambda \in \Sigma$ *is called a "blank symbol"*

$Q$ *is a finite set, called "set of states"*

$q_i$ *is the initial state*

$q_f$ *is the final state*

$\delta : \Sigma \times Q \to (\Sigma \times Q \times \{L, R\})^{|Q| \cdot |\Sigma| \cdot 2}$ *is a transition function*

$q_q \in Q$ *is a special query state*

$q_a \in Q$ *is a special post-query state*

*and the corresponding finite transition matrix is unitary for all lengths of input. When the machine enters the state $q_q$, let the contents of the tape from the current position to the first blank symbol be $|x, b\rangle$, where $b$ is a single qubit. Then the oracle transforms the state to $|x, b \oplus I(x)\rangle$, where $I(x)$ is the indicator of $x$ and the machine enters the post-query state $q_a$.*

### 2.2.2 The quantumness of a quantum Turing machine

The natural question is now: "Where is the quantumness of a quantum Turing machine hidden, if there ever is any?" In other words, what property makes the quantum Turing machines possibly more powerful?

As for example [5] suggest, it is the following feature of quantum Turing machines: Two different paths may interfere both constructively (if their amplitudes are equal), or destructively, if their amplitudes are opposite. Here we justify it by the following by the following observation:

**Observation 2.4.** *If all amplitudes in a quantum Turing machine $M$ are real positive, then there exists a Turing machine accepting the same language.*

*Proof.* From the assumption about amplitudes, it follows that in the transition matrix of $M$, there are only real positive numbers. That means that as the entries there are only 1s and 0s. Let us assume the opposite and have a row $r$ with two or more nonzero entries, which can be, without loss of generality, symbolically written as

$$r = \sum_{i=1}^{k} a_i e_i \tag{2.1}$$

where $k \geq 2$ and $e_i$ are base vectors[3] and $a_i$ are real positive coefficients. Then there must be at least one more row $r'$ that has a nonzero coefficient, let us say $b_1$, on the first place, because otherwise the first column could not have a unit norm. This implies that $rr' \geq a_1 b_1 > 0$ which violates the unitarity condition so we have arrived at a contradiction. Thus, the transition matrix is a permutation matrix and the machine $M$ now works as a classical Turing machine, since all squares of amplitudes are equal to 1. □

---

[3] of the same base where the matrix is written

# Chapter 3

# Quantum circuits

In this chapter we will define quantum circuits and show their strengths or weaknesses comparing to classical Turing machines. We will not talk about quantum Turing machines, this is left to chapter 4. We will however not introduce the concept of classical circuits, which would be at the first sight a more natural model to compare the quantum circuits with. The reason is that the quantum circuits have some important intristic properties different form those of classical circuits and so comparing these two is unnecessarily complicated. Furthermore, we will only introduce uniform complexity classes based on circuits, as is the class $BQP$.

**Definition 3.1 (Quantum circuit).** *Let us have a two-dimensional Hilbert space $H$. Quantum circuit of width $n$ in that space is a unitary operator $U$ of size $2^n \times 2^n$ working on joint systems composed of $n$ qubits from $H$, which are initially in some base state of $H^n$. $U$ thus realizes some function $f : (\text{base of } H^n) \to H^n$. If we identify the base vectors of $H$, $|0\rangle$ and $|1\rangle$, with logical 0 and 1, and the output with some probability distribution by squaring the norms of amplitudes, we may say that for each bit string of length $n$, the circuit outputs some probability distribution over all bit strings. We may also treat some subsystems as auxiliary, in that case we forget about them when evaluating the output.*

A circuit of width $n$ is visualized in Figure 3.1. The horizontal lines are called *wires*. The joint system we work with and also the respective subsystems are called a *quantum register*.

The unit $M$, a so called *measurement operator* is there to emphasize we look at the output as at a probability distribution and forget the complex values of amplitudes. This is in accordance with quantum mechanics which tells us that after being observed, a quantum system collapses to one of the basis vectors with a probability corresponding to its amplitude.[1]

The computational power of classical computers is expressed by Church-Turing thesis: "Any computational device can be simulated by a Turing machine." It also holds that

---
[1]Here we silently assume the measurement is done in the same basis as we work in.

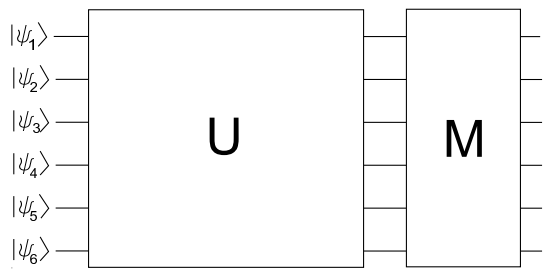

Figure 3.1: *A visualization of a quantum computer. $U$ is a unitary matrix and $M$ is a measurement operator. $U$ is applied onto $|\psi_1\rangle \otimes |\psi_1\rangle \otimes |\psi_2\rangle \otimes \ldots \otimes |\psi_6\rangle$ and to the result a measuring operator $M$ is applied.*

**Lemma 3.1 (Power of quantum circuits).** *For any Turing machine $T$ that halts for all inputs and any bit string $I$, there exists a unitary matrix $U$, whose action result on the $I$ encoded into a quantum register leads to the same output as of $T$ on $I$.*

*Proof.* Each Turing machine $M$ realizes some computable function $x \to M(x)$. Each such input and output can be encoded into bits and subsequently into qubits. A matrix $U$ realizing transformation $(x, 0) \to (x, M(x))$ is certainly unitary, since we have that

$$(x, 0) \perp (y, 0) \iff x \perp y \iff (x, M(x)) \perp (y, M(y))$$

This means that we do not start the algorithm only with a register containing $x$, but add another register containing only zeros, which are usually called "padding" zeros. If we did not use them, the transform realizing $x \to M(x)$ would not generally be unitary. $\square$

We see that computational power of a quantum circuit is superior or equal to the one of a classical Turing machine.

## 3.1 Universal sets of gates

When working with quantum circuits, we arrive at a need to define a complexity measure. With $n$ wires, the matrix $U$ in Figure 3.1 has a size of $2^n \times 2^n$. Thus, the size of the matrix can not be a good measure of complexity, since all algorithms would have exponential complexity. Furthermore, if we have $U = 1$ we are practically doing nothing and the complexity should be 0, while the size of that matrix remains exponential. Thus, the complexity measure should somehow reflect how complicated the structure of the matrix is. Also, it should hold that matrices of a low complexity can be practically realized easier than those with a higher complexity. Furthermore, if we work with fewer qubits, the corresponding complexity should lower. Thus,

**Requirement 3.1.1.** *If $U = U_1 U_2$ then its complexity measure $C$ should be $C(U) \leq C(U_1) + C(U_2)$*
*If $U = C(U_1 \otimes U_2)$ then its complexity measure should be $C(U) \leq C(U_1) + C(U_2)$.*

The requirement is illustrated in figure 3.2. Now, if we choose some set of gates, called

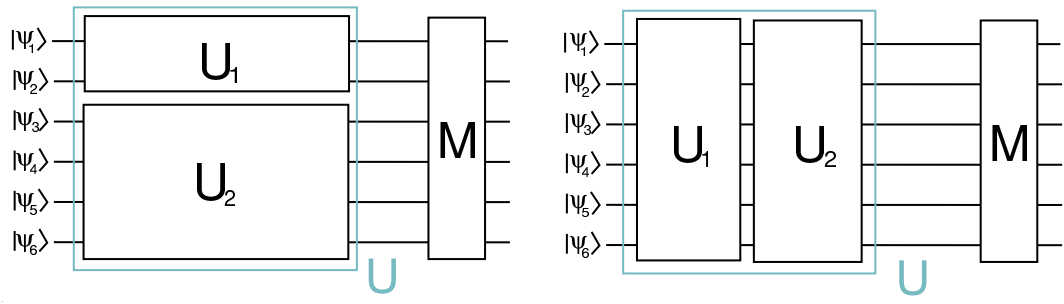

Figure 3.2: A composite unitary matrix. The left picture shows $U = U_1 \otimes U_2$ and the right $U = U_1 U_2$. It should be pointed out here again, that the size of a gate working on $n$ qubits is always $2^n$ since it transforms an $n$-qubit string into another, of which there are $2^n$ possible mutually orthogonal combinations. And since its unitary, such a matrix is always square-shaped. Thus, the width of the gates in the figure is irrelevant. On the left, we have two gates of sizes $2^2$ and $2^4$ respectively, while on the right, we have two gates of size $2^6$.

*elementary set*, and set the complexity of gates from that set to 1, we can construct upper bounds of the complexity using the Requirement 3.1.1. To proceed, we introduce some definitions.

**Definition 3.2 (Universal set).** *A set of quantum gates is said to be* universal *for quantum computation if any unitary operator can be expressed as a circuit involving only gates from that set.*

Many times, when we are dealing with probabilistic quantum algorithms, we may allow for small differences in the gates. We may suffice with a gate that is almost equal to the gate we want.

**Definition 3.3 (Approximately universal set).** *A set of quantum gates is said to be* approximately universal *for quantum computation if any unitary operator can be approximated with arbitrary precision by a circuit involving only those gates.* [2]

To proceed further, we will define some important gates.

## 3.2 Some important gates

It was shown that any unitary gate can be built up from one and two-qubit gates. This fact is important for an eventual quantum computer, since such gates are easier to prepare and manipulate. Here we will define some of them, which we will need later.

**One qubit gates**
The *Hadamard gate* is defined as

$$H = \frac{1}{\sqrt{2}} \begin{pmatrix} 1 & 1 \\ 1 & -1 \end{pmatrix}$$

It acts on a ket vector $a|0\rangle + b|1\rangle$ as

$$H(a|0\rangle + b|1\rangle) = \frac{1}{\sqrt{2}} \begin{pmatrix} 1 & 1 \\ 1 & -1 \end{pmatrix} \left( a \begin{pmatrix} 1 \\ 0 \end{pmatrix} + b \begin{pmatrix} 0 \\ 1 \end{pmatrix} \right) =$$

$$= \frac{1}{\sqrt{2}} \left( (a+b) \begin{pmatrix} 1 \\ 0 \end{pmatrix} + (a-b) \begin{pmatrix} 0 \\ 1 \end{pmatrix} \right) = \frac{1}{\sqrt{2}} \left( (a+b)|0\rangle + (a-b)|1\rangle \right)$$

The Hadamard gate is of a great importance, since a tensor product of $n$ Hadamard gates can be used to prepare $n$-qubit quantum register in a linear superposition of all $2^n$ base states $|j\rangle$:

$$H^n \stackrel{\text{DEF}}{=} \left( \bigotimes_{i=0}^{n-1} H \right) |0^n\rangle = \bigotimes_{i=0}^{n-1} H|0\rangle = \frac{1}{\sqrt{2^n}} \bigotimes_{i=0}^{n-1} (|0\rangle + |1\rangle) = \frac{1}{\sqrt{2^n}} \sum_{j=0}^{2^n} |j\rangle$$

Notice that we have needed only $n$ Hadamard gates to prepare a superposition of exponentially many -$2^n$ - elements. This feature is of a great use in quantum algorithms discussed later on.

A *phase shift gate* is for any $\theta \in \mathbb{R}$ defined as

$$P(\theta) = \begin{pmatrix} 1 & 0 \\ 0 & e^{i\theta} \end{pmatrix}$$

and acts acts

$$P(\theta)(a|0\rangle + b|1\rangle) = \begin{pmatrix} 1 & 0 \\ 0 & e^{i\theta} \end{pmatrix} \left( a \begin{pmatrix} 1 \\ 0 \end{pmatrix} + b \begin{pmatrix} 0 \\ 1 \end{pmatrix} \right) =$$

$$= \left( a \begin{pmatrix} 1 \\ 0 \end{pmatrix} + be^{i\theta} \begin{pmatrix} 0 \\ 1 \end{pmatrix} \right).$$

---

[2]Since we say "arbitrary precision" here, we do not bother about the metric we use to define the distance between the matrices and thus the precision. It can be just any metric, if the word arbitrary means "any nonzero value in the metrics range".

**Two-qubit gates**
*CNOT gate*

$$CNOT = \begin{pmatrix} 1 & 0 & 0 & 0 \\ 0 & 1 & 0 & 0 \\ 0 & 0 & 0 & 1 \\ 0 & 0 & 1 & 0 \end{pmatrix}$$

It acts on a two-qubit state $a|00\rangle + b|01\rangle + c|10\rangle + d|11\rangle$ as

$$CNOT\left(a|00\rangle + b|01\rangle + c|10\rangle + d|11\rangle\right) =$$

$$= \begin{pmatrix} 1 & 0 & 0 & 0 \\ 0 & 1 & 0 & 0 \\ 0 & 0 & 0 & 1 \\ 0 & 0 & 1 & 0 \end{pmatrix} \left( a\begin{pmatrix}1\\0\\0\\0\end{pmatrix} + b\begin{pmatrix}0\\1\\0\\0\end{pmatrix} + c\begin{pmatrix}0\\0\\1\\0\end{pmatrix} + d\begin{pmatrix}0\\0\\0\\1\end{pmatrix} \right) =$$

$$= a\begin{pmatrix}1\\0\\0\\0\end{pmatrix} + b\begin{pmatrix}0\\1\\0\\0\end{pmatrix} + d\begin{pmatrix}0\\0\\1\\0\end{pmatrix} + c\begin{pmatrix}0\\0\\0\\1\end{pmatrix}$$

An important property of CNOT gate is that it is inseparable, e.g. it cannot be written as a product of two one-qubit gates.

Now, we are ready to introduce some universal sets. In [3], elementary sets are presented. We will suffice with the following group:

**Lemma 3.2.** *The following sets of gates are universal:*

- *For a constant $k$, set of all gates of size $k \times k$.*

- *All one-qubit gates and CNOT*

- *$\{P(\theta), CNOT, H\}$*

One of the most widely used universal sets is *CNOT* together with all one qubit gates. We often treat this set as elementary and use it to measure complexity. In this case, it has no sense to distinguish subpolynomial complexities. Nevertheless, many different small subsets of this elementary set, like $\{CNOT, H, Phase(\frac{\pi}{8})\}$ are approximately universal. Here, when having only three elements, we can measure complexity precisely. The universal set is on the other side more comfortable to use when working with complexity classes up to polynomial reductions.

It turns out that the complexity of a general unitary transform is most likely exponential. In other words, a general unitary gate working on $n$ quantum registers can not be composed from less then $\Omega(4^{n-2})$ two-qubit gates. The argument is simple:

**Lemma 3.3.** *For each $n$ there exists a gate of size $2^n \times 2^n$ working on $n$ qubits, such that complexity of that gate is $\Theta(4^{n-2})$.*

*Proof.* A unitary gate of size $2^n \times 2^n$ has $2^{2n} = 4^n$ free angle parameters. A two qubit gate has $2^{2\cdot 2} = 4^2$ parameters. A number of free parameters does not change when doing a tensor product with 1. To collect $4^n$ free parameters, we need at least $\frac{4^n}{4^2} = 4^{n-2}$ two qubit gates. □

The lemma suggests that most of quantum algorithms will have too high a complexity to be of any use. Due to another result in [3], any two qubit gate can be expressed by four one qubit gates and two *CNOT* gates, we don't have to look at their concrete form of one-qubit and two-qubit gates we use, but only at their dimension. The following lemma that appeared at [3] shows, that the lower bound is close to the upper bound:

**Lemma 3.4.** *Each unitary matrix of size $2^n \times 2^n$ can be decomposed to at most $2^{O(n)}$ one qubit and* CNOT *gates.*

Using the notation from the theory of classical circuits, we rather refer to a size of a circuit instead of complexity:

**Definition 3.4 (Size of a circuit).** *A size of a circuit C with respect to an universal (resp. approximately universal) set of gates is the complexity with respect to that set.*

If we are talking about sizes up to a polynomial, we may omit to name a concrete set of gates, thanks to lemma 3.4, provided we have in mind that we should refer only to universal sets of gates of bounded size.

Now, we will define how can quantum circuits accept languages. The function $a(x)$ has the same role as in the definition 2.16. It tells us which configuration is the accepting one for each $x$. Each circuit obviously works only on inputs with some predefined size.

**Definition 3.5 (Acceptation of a language).** *We say that a quantum circuit C of width n working on Hilbert space H accepts language L iff there is a function $a(x) : (base\ of\ H) \to base\ of\ H$ such that for each x of length n it outputs a state $\sum_i a_i |e_i\rangle$ such that*

*If $x \in L$ then $\sum_i a_i |\langle a(x)|e_i\rangle|^2 > \frac{2}{3}$*

*If $x \notin L$ then $\sum_i a_i |\langle a(x)|e_i\rangle|^2 \leq \frac{1}{3}$*

*where $\{e_i\}$ is a base of H.*

## 3.3 Entanglement

The phenomenon of entanglement has been extensively treated by many scientists (see for example [11], [2]) and also mystified in many articles. It arises as follows: A quantum system composed of $n$ qubits lives in a Hilbert space of dimension $2^n$, generated by for example the following base

$$\bigcup_{i=0}^{2^{n-1}} \{|i\rangle\} \tag{3.1}$$

On the other side, a linear space generated by vectors of length $n$ with complex coefficients has dimension $n$ and can be injectively mapped onto a proper subset of that Hilbert space. The mapping works as

$$(c_1, c_2, \ldots, c_n) \leftrightarrow \bigotimes_{i=1}^{n} \left( \sqrt{1 - |c_i|^2}|0\rangle + c_i|1\rangle \right) \tag{3.2}$$

and is a natural mapping of complex vectors onto qubit states. The state of the $i-$th qubit straightforwardly corresponds to the $i-$th component of the complex vector. Such quantum states are called *disentangled* or *separable states*. The complement of that subset to the entire Hilbert space is filled by *entangled* states. This is where the mysterious spirit of entangled states comes from. They don't have their counterparts in a complex space of vectors of length $n$ and are thus maybe a bit counterintuitive. It can be seen that separable states generate the whole Hilbert space. It follows that any vector from the Hilbert space can be expressed as a sum of separable states. Before describing its magical features, we will state a simple observation.

**Observation 3.5 (The role of entanglement).** *Without entanglement, quantum computer is polynomially equivalent to Turing machine.*

*Proof.* In a quantum computer with $n$-qubit register without entanglement, at every stage of computation, the register is in a separable state. Thus, at every stage, the register's state can be described by a complex vector of length $n$. For each two vectors, there exists a general matrix realizing a transformation from one to another. Thus, the whole computation can be seen as a simple matrix multiplication. Each Turing machine can be written as a matrix multiplication and each multiplication can be realized on a Turing machine, with only a polynomial slowdown. □

By further examination of properties of separable states, we may state a stronger version

**Observation 3.6.** *Without entanglement, quantum computer can be simulated on a Turing machine in linear time.*

*Proof.* Because the gates in the quantum computer (again with $n$ qubits) can not lead to inseparable states, the gates have to be separable itself. That means the circuit is composed of only one qubit gates. They can be at most $n$. If they are more then 1 on a single wire, they can be multiplied up. Thus, the complexity of simulating the circuit is $n$ times the complexity of multiplying a matrix of size $2 \times 2$ by a vector, which is constant. □

However, one should note that a quantum circuit with $n$ gates has complexity at least $n$. Therefore, the complexities of both the machines are asymptotically equal.

We have seen that for a quantum circuit to ever perform asymptotically better than classical computers, entanglement is crucial. If it ever happens that there is no entanglement in a quantum algorithm and still it performs more than linearly better than its classical counterpart, it means that the latter can be improved.

## 3.4 Quantum query model

As we let quantum Turing machines access additional information via oracles, we will also allow the circuits to access some kind of additional information. Here, in the model of quantum circuits, the oracle is picturized as a special type of a gate, called "black box" to emphasize we can not see its internal structure. It acts on registers of the form $|x, b\rangle$ as

$$|x, b\rangle \underrightarrow{O} |x, y \oplus I(x)\rangle,$$

where $b$ is a single qubit and $I(x)$ is the indicator of $x$. This unit is thus effectively equivalent to the oracle unit used in oracle quantum Turing machines. As before, we need to assume one more thing comparing to classical black boxes: The quantum black-box needs to support superposition, e.g. to behave as a linear operator. The black-box model has turned out to be very fruitful for various results, see for example [6].

# Chapter 4

# Putting it together

In this chapter, we will show the relation between QTMs and quantum circuits. We will review the definitions of conventional complexity classes and define the class BQP.

## 4.1 Relation between QTMs and Quantum circuits

To ever talk about the relation between quantum Turing machine and quantum circuits, we have to define what simulation of respective things mean.

**Definition 4.1 (Simulation).** *We say that a quantum Turing machine Q simulates a quantum circuit C on an input I, if Q, fed with I, gives as output probability distribution identical to the one that C gives.*
*We say that a circuit C with input I simulates a quantum Turing machine Q on that input, if they both on that input output the same probability distribution.*

To show the relation between quantum circuits and quantum Turing machines, we introduce the following observation:

**Observation 4.1.** *For each constant integer $k$ and unitary matrix $U$ of size $2^k \times 2^k$ working on $k$ qubits, there exists a quantum Turing machine that simulates circuit containing only $U$.*

*Proof.* That matrix can be thought as a function on an appropriate Hilbert space. For each constant $k$, such function can be realized on some quantum Turing machine. □

It is necessary that $k$ be a constant for the observation to hold, since for the simulation, the quantum Turing machine must have $|\Sigma| + |Q|$ of size at least $log(2^k) = k$.

**Lemma 4.2.** *For each circuit of size $n$ there exists a quantum Turing machinewith time complexity $T(n)$ simulating that circuit such that $T(n) = O(n)$.*

*Proof.* Each gate of size $2^k$ can be simulated on a quantum Turing machineusing at most $2^{O(k)}$ steps. Thus, if the maximal size of gates in the universal set is $m$, the simulation takes time at most $2^{O(m)}n = O(n)$ steps. □

The converse relation between quantum circuit and a quantum Turing machine was proved at [16]. Here we present a slightly simplified version:

**Lemma 4.3.** *For each integer $n$ and each quantum Turing machine $Q$ of time complexity $T$, there exists a circuit of $poly(n, T)$ elementary gates that simulates $Q$ on any input of length $n$.*

*Proof.* For each of $T$ steps of Q, a separate circuit is constructed. The machine cannot go further than to distance of $T$ from the beginning position of the tape, so we cut off everything except the middle $2T + 1$ cells. For each cell, we add $l = (1 + \lceil log(|Q| + 1) \rceil + \lceil log|\Sigma| \rceil)$ wires to the resulting circuit. These wires encode:

- the current state: $\lceil log(|Q| + 1) \rceil$ wires, where the 1 stays for a new state $s$ (if the machine never reaches there)

- the symbol on that cell: $\lceil log|\Sigma| \rceil$ wires

- a bit indicating whether the head is on respective cell 1=yes

Then for each step and cell, we take the cell the head is just on and the two neighbors, together tree cells. This suffices, since in one step, the head cannot escape from them. Furthermore, we know that without loss of generality, we may not let the machine stay at the same state after a transition. Now we want to have a unitary matrix $U$ that would transform the state on the $3l$ wires according to how the transition function $\delta$ would do. Formally, we want that

$$U|(n,a_l,0)(q,a,1)(n,a_r,0)\rangle = \sum_{a',q'} \delta(q,a,q',a',L)|(q',a_l,1)(n,a',0)(n,a_r,0)\rangle + \\ \delta(q,a,q',a',R)|(n,a_l,0)(n,a',0)(q',a_r,1)\rangle \quad (4.1)$$

Luckily, it holds that for all distinct pairs of $q, a, a_r, a_l$, the vectors $U|(n,a_l,0)(q,a,1)(n,a_r,0)\rangle$ are mutually orthogonal:

$$\langle (s,a_{l1},0)(q_1,a_1,1)(s,a_{r1},0)|U^+U|(s,a_{l2},0)(q_2,a_2,1)(s,a_{r2},0)\rangle = \\ \Big( \sum_{a'_1,q'_1} \delta(q_1,a_1,q'_1,a'_1,L)\langle (q'_1,a_{l1},1)(s,a'_1,0)(s,a_{r1},0)| + \\ \delta(q_1,a_1,q'_1,a'_1,R)\langle (s,a_{l1},0)(s,a'_1,0)(q'_1,a_{r1},1)| \Big) \\ \Big( \sum_{a'_2,q'_2} \delta(q_2,a_2,q'_2,a'_2,L)|(q'_2,a_{l2},1)(s,a'_2,0)(s,a_{r2},0)\rangle + \\ \delta(q_2,a_2,q'_2,a'_2,R)|(s,a_{l2},0)(s,a'_2,0)(q'_2,a_{r2},1)\rangle \Big) = \\ \sum_{a'_1,q'_1,a'_2,q'_2} \delta(q_1,a_1,q'_1,a'_1,L)\delta(q_2,a_2,q'_2,a'_2,L) \\ \langle (q'_1,a_{l1},1)(s,a'_1,0)(s,a_{r1},0)|(q'_2,a_{l2},1)(s,a'_2,0)(s,a_{r2},0)\rangle + \\ \delta(q_1,a_1,q'_1,a'_1,R)\delta(q_2,a_2,q'_2,a'_2,R) \\ \langle (s,a_{l1},0)(s,a'_1,0)(q'_1,a_{r1},1)|(s,a_{l2},0)(s,a'_2,0)(q'_2,a_{r2},1)\rangle = \\ \delta_{a_{l1}a_{l2}}\delta_{a_{r1},a_{r2}} \Big( \sum_{a'_1,q'_1} \delta(q_1,a_1,q'_1,a'_1,L)\delta(q_2,a_2,q'_1,a'_1,L) + \delta(q_1,a_1,q'_1,a'_1,R)\delta(q_2,a_2,q'_1,a'_1,R) \Big) \\ = 0 \quad (4.2)$$

Where the last equality follows from the unitarity condition of the quantum Turing machine, because in the transition matrix all pairs of rows are mutually orthogonal. The rest of base vectors can be thus set so that $U$ is a unitary matrix. We add such matrices to each 3 consecutive cells. The order does not matter thanks to the definition of $U$. This simulates one step of Q. To simulate $T$ steps, we just chain $T$ such circuits.

We have together used $O(T(2T+1))$ matrices of size $2^{3l} \times 2^{3l}$ and $2O(l(2T+1))$ wires. From Lemma 3.4, we know that such matrices can be realized by $2^{O(l)}$ elementary gates. In sum, we used $(T(2T+1))2^{O(l)} = T^2 2^{O(l)}$ elementary gates. □

## 4.2 Complexity classes

### 4.2.1 Classical classes

For sake of completeness, we briefly review classical complexity classes definitions.

**Definition 4.2 (Classical polynomial time complexity classes).** *A language $L$ is in class $C$ if there exists a polynomial $p(n)$ and a Turing machine $M$ with time complexity $p(n)$ such that*

| complexity class $C$ | $P$ | $NP$ | $PP$ | $BPP$ |
|---|---|---|---|---|
| For $x \in L$, $P_{y:|y|=p(n)}(M(x,y) \, accepts)$ | $=1$ | $>0$ | $>\frac{1}{2}$ | $\geq \frac{2}{3}$ |
| For $x \notin L$, $P_{y:|y|=p(n)}(M(x,y) \, accepts)$ | $=0$ | $=0$ | $\leq \frac{1}{2}$ | $\leq \frac{1}{3}$ |

*where exactly one of the columns applies.*

The abbreviations stand for polynomial, nondeterministic polynomial, probabilistic polynomial, and bounded probability polynomial respectively.

All of these classes are $\Sigma_2$ definable, which means that is one $\exists$ and one $\forall$ quantifier before a general predicate computable in the class $P$. That is considered to be a condition of a robust definition.

In the future, we will need different numbers in the definition of the class $BPP$ then $\frac{1}{3}$ and $\frac{2}{3}$. Therefore, we will introduce the following two lemmas. The first is a wellknown bound on the tail of the distribution of a sum of random variables and can be found for example in [15]. The second, amplification of the accepting probability, can be found in many complexity books, for example in [12].

**Lemma 4.4 (Chernoff bounds).** *Suppose $x_1, x_2, \ldots x_n$ are chosen independently from a fixed distribution. Let the mean value of $x$ be $\mu \stackrel{DEF}{=} E(x)$. Then for any positive $\lambda$ the following holds:*

$$P\left(\left|\frac{\sum_{i=1}^{N} x_i}{N} - \mu\right| \geq \lambda\right) \leq e^{-\lambda^2 \frac{N}{2}}$$

**Lemma 4.5 (BPP amplification).** *The numbers $\frac{1}{3}$ and $\frac{2}{3}$ in the definition of the class $BPP$ may be equivalently changed to any $a$ and $b$ such that $a \geq b + \epsilon$ for some $\epsilon > 0$.*

*Proof.* Let us have an algorithm that accepts $x \in L$ with probability $a$ and $x \notin L$ with probability $b$. We run the algorithm $k(n)$ times for $k(n)$ being an even polynomial and then accept if and only iff at least $\frac{a+b}{2}$ of the runs accepted. Now we define new quantities $z_i$ in the following way:

$$z_i = \begin{cases} 1 & \text{if i-th run accepted} \\ 0 & \text{otherwise} \end{cases}$$

For, $x \in L$, we have that $E(z) \geq 1 \cdot a + 0 \cdot (1-a) = a$ and
for $x \notin L$, we have that $E(z) \leq 1 \cdot b + 0 \cdot (1-b) = b$. Now we are ready to use Lemma 4.4. We have

$$P(x \in L \text{ is accepted}) = P\left(\sum_{i=1}^{k} x_i \geq k\frac{a+b}{2}\right)$$

$$\geq P\left(\left|\frac{\sum_{i=1}^{k} x_i}{k} - a\right| \leq \frac{a-b}{2}\right) = 1 - P\left(\left|\frac{\sum_{i=1}^{k} x_i}{k} - a\right| \geq \frac{a-b}{2}\right) \geq 1 - e^{-\left(\frac{a-b}{2}\right)^2 \frac{k}{2}}$$

and

$$P(x \notin L \text{ is accepted}) = P\left(\sum_{i=1}^{k} x_i \geq k\frac{a+b}{2}\right) \leq P\left(\left|\frac{\sum_{i=1}^{k} x_i}{k} - b\right| \geq \frac{a-b}{2}\right) \leq e^{-\left(\frac{a-b}{2}\right)^2 \frac{k}{2}}$$

Now it suffices to define $k(n)$ big enough so that $e^{-\left(\frac{a-b}{2}\right)^2 \frac{k}{2}} \leq \frac{1}{3}$ and $1 - e^{-\left(\frac{a-b}{2}\right)^2 \frac{k}{2}} \geq \frac{2}{3}$. Conversely, if we start with the original numbers $\frac{2}{3}$ and $\frac{1}{3}$, we can reach any other pair $a$ and $b$ satisfying the assumption $a \geq b + \epsilon$ for some $\epsilon > 0$ by setting the polynomial $k(n)$ big enough so that $e^{-\left(\frac{1}{6}\right)^2 \frac{k}{2}} \leq b$ and $1 - e^{-\left(\frac{1}{6}\right)^2 \frac{k}{2}} \geq a$. □

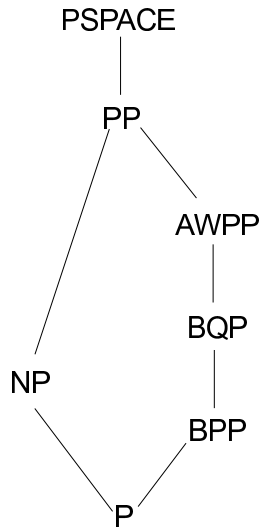

Figure 4.1: *Hierarchy of classes. For each pair connected by a line, the class that stays upper contains the lower one.*

**Definition 4.3 (PSPACE).** *A language $L$ is in $PSPACE$ if there exists a polynomial $p(n)$ and a Turing machine $M$ with space complexity $p(n)$ such that $L$ is accepted on $M$.*

### 4.2.2 BQP class

**Definition 4.4 (BQP).** *A language $L$ is in $BQP$ if there exists a polynomial $p(n)$ such that $L$ is accepted by some quantum Turing machine with time complexity $p(n)$.*

Thanks to Lemma 4.3 and Observation 4.1, we may equivalently define $BQP$ as

**Definition 4.5 (BQP).** *A language $L$ is in $BQP$ if there exists a function $f(n)$ and polynomials $p(n), q(n)$ such that for each $n$, output of $f(n)$ is a circuit $C$ of width $n$ and size $p(n)$ such that the language $L_n \equiv \{x \in L : |x| = n\}$ is accepted by $C$ and the running time of $f(n)$ is at most $q(n)$.*

The position of the class $BQP$ in the hierarchy of other classes is shown in figure 4.1.

# Chapter 5

# New quantum classes

In this chapter we will define new complexity classes. The goal will be to provide an easy to handle hierarchy of simple classes in which the class $BQP$ is contained, and take profit of it to show some theorems. We will begin with the definition of $BQP$, which we reformulated in a matrix fashion.

## 5.1 The class $BQP$ revisited

**Definition 5.1** ($BQP$). *A language $L$ is in $BQP$ if there is a function $f : \mathbb{N}-> $ (set of all quantum circuits) computable in polynomial time with respect to its argument, such that for each $l$, its output is a polynomial size quantum circuit $T$ and two functions $c_I(x)$ and $c_A(x)$ computable in polynomial time such that for each $x$ of length $l$:*

*For $x \in L$, we have $\left|\langle c_A(x)|T|c_I(x)\rangle\right|^2 \geq \frac{2}{3}$.*

*For $x \notin L$, we have $\left|\langle c_A(x)|T|c_I(x)\rangle\right|^2 \leq \frac{1}{3}$.*

In this and all the definitions that follow, we do not require that the functions $c_I, c_A$ have the same range and domain. We for example allow for $|c_I(x)\rangle = |x, x\rangle$. The function $c_I(x)$ corresponds to some transformation of the input, for example adding 'padding' zeros or whatever. It can be arbitrary function as long as it is computable in polynomial time. The function $c_A(x)$ expresses the fact that the accepting state of the register typically depends on the input. The only thing we require is that given $x$, we can compute in a polynomial time for which state we are looking for. The equivalence of Definition 5.1 and Definition 4.5 is due to Definition 3.5.

In the future, we will frequently need a higher number than the $\frac{2}{3}$ in the definition. For this purpose, we have to prove the following three lemmas. Although its is done compactly in [5], we split it into three lemmas here, because later we will need those parts separately.

**Lemma 5.1.** *Let us have a language $L$, a polynomial size circuit $T$ and a function $c_I(x)$ computable in polynomial time. Let us have a function $a(x, y)$ which decides, for a given input $x$, if $y$ is an accepting configuration or not. If so, it outputs $1$, otherwise it outputs $0$. Let us also assume the following:*

*For $x \in L$, we have $\sum_{y:a(x,y)=1}\left|\langle y|T|c_I(x)\rangle\right|^2 \geq \frac{2}{3}$.*

*For $x \notin L$, we have $\sum_{y:a(x,y)=1} = \left|\langle y|T|c_I(x)\rangle\right|^2 \leq \frac{1}{3}$.*

*We can then equivalently amplify the numbers $\frac{2}{3}$ and $\frac{1}{3}$ into $1-e^{q(n)}$ and $e^{q(n)}$ for any positive polynomial $q(n)$.*

*Proof.* This can be done by replicating the circuit $k(n)$ for some polynomial $k$. As in Lemma 4.5, we will do a majority vote. The new function $a'(x, y_1 y_2 \ldots y_{k(n)})$ will check that for at least half of the $y$s there is $a(x,y) = 1$. We have

$$T' \equiv \bigotimes_{i=1}^{k} M$$

and

$$a'(x, y_1 y_2 \ldots y_{k(n)}) \stackrel{\text{DEF}}{=} \begin{cases} 1 & \text{if at least a half of the } y \text{ equals } c_A(x) \\ 0 & \text{otherwise} \end{cases}$$

We define $z_i$ as in Lemma 4.5 and proceed analogously. Then we have

for $x \in L$:

$$\sum_{y: a'(x, y_1 y_2 \ldots y_{k(n)}) = 1} \left| \langle y | T'^{t(n)} | c_I(x) \rangle \right|^2 = P\left( \sum_{i=1}^{k} z_i \geq \frac{k(n)}{2} \right) \geq 1 - e^{-(\frac{1}{6})^2 \frac{k}{2}}$$

for $x \notin L$:

$$\sum_{y: a'(x, y_1 y_2 \ldots y_{k(n)}) = 1} \left| \langle y | T'^{t(n)} | c_I(x) \rangle \right|^2 = P\left( \sum_{i=1}^{k} z_i \leq \frac{k(n)}{2} \right) \leq e^{-(\frac{1}{6})^2 \frac{k}{2}}$$

It now suffices to set $k(n) = 2 \cdot 36 q(n)$. □

The following lemma [5] shows that for the class $BQP$, we may equivalently require that there be more accepting configurations:

**Lemma 5.2.** *Let us have a language $L$, a polynomial size circuit $T$ and a function $c_I(x)$ computable in polynomial time. Let us have a function $a(x,y)$ which decides, for a given input $x$, if $y$ is an accepting configuration or not. If so, it outputs $1$, otherwise it outputs $0$. Let us also assume the following:*

*For $x \in L$, we have $\sum_{y: a(x,y)=1} \left| \langle y | T | c_I(x) \rangle \right|^2 \geq \frac{2}{3}$.*

*For $x \notin L$, we have $\sum_{y: a(x,y)=1} = \left| \langle y | T | c_I(x) \rangle \right|^2 \leq \frac{1}{3}$.*

*We can then build a quantum circuit accepting the same language showing that $L \in BQP$.*

*Proof.* Using Lemma 5.1, we know that we may assume the following:

For $x \in L$, we have $\sum_{y: a(x,y)=1} \left| \langle y | T | c_I(x) \rangle \right|^2 \geq \frac{8}{9}$.

For $x \notin L$, we have $\sum_{y: a(x,y)=1} = \left| \langle y | T | c_I(x) \rangle \right|^2 \leq \frac{1}{9}$.

Now, we replicate the input and add another wire in state $|0\rangle$. To the end of the resulting circuit, we add another polynomial size circuit that would computes

$$f(y, x, 0) \stackrel{\text{DEF}}{=} \begin{cases} |y, x, 1\rangle & \text{if } a(x,y) = 1 \\ |y, x, 0\rangle & \text{if } a(x,y) = 0 \end{cases}$$

Such circuit certainly exists since $a$ is computable in polytime and $f$ is injective. Then, we add $T^{-1}$ working on the first register, to transform $(x, y, i) \to (x, x, i)$. We then set

$$c'_I(x) \stackrel{\text{DEF}}{=} |x, x, 0\rangle$$
$$c'_A(x) \stackrel{\text{DEF}}{=} |x, x, 1\rangle$$

The resulting polynomial circuit will be denoted by $T'$.

Now, we look at $\left|\langle c'_A(x)|T'|c'_I(x)\rangle\right|$. We have

$$\langle c'_A(x)|T'|c'_I(x)\rangle = \langle x,x,1|T^{-1}fT|x,x,0\rangle =$$

$$\langle x,x,1|T^{-1}f \sum_{y:a(x,y)=1} |y,x,0\rangle\langle y,x,0|T|x,x,0\rangle +$$

$$\langle x,x,1|T^{-1}f \sum_{y:a(x,y)=0} |y,x,0\rangle\langle y,x,0|T|x,x,0\rangle$$

$$= \sum_{y:a(x,y)=1} \langle x,x,1|T^\dagger|y,x,1\rangle\langle y,x,0|T|x,x,0\rangle +$$

$$\sum_{y:a(x,y)=0} \langle x,x,1|T^\dagger|y,x,0\rangle\langle y,x,0|T|x,x,0\rangle$$

$$= \sum_{y:a(x,y)=1} \langle y,x,1|T^*|x,x,1\rangle\langle y,x,0|T|x,x,0\rangle +$$

$$\sum_{y:a(x,y)=0} \langle y,x,0|T^*|x,x,1\rangle\langle y,x,0|T|x,x,0\rangle$$

$$= \sum_{y:a(x,y)=1} |\langle y|T|x\rangle|^2$$

Then for $x \in L$, we have that

$$\left|\langle c'_A(x)|T'|c'_I(x)\rangle\right|^2 = \left|\sum_{y:a(x,y)=1} \left|\langle y|T|x\rangle\right|^2\right|^2 \geq 0.9^2 \geq \frac{2}{3}$$

and for $x \notin L$

$$\left|\langle c'_A(x)|T'|c'_I(x)\rangle\right|^2 = \left|\sum_{y:a(x,y)=1} \left|\langle y|T|x\rangle\right|^2\right|^2 \leq 0.1^2 \leq \frac{1}{3}$$

□

**Lemma 5.3 (Probability amplification).** *The numbers $\frac{2}{3}, \frac{1}{3}$ in Definition 5.1 can be amplified to $1 - e^{-q(n)}$, $e^{-q(n)}$ for an arbitrary positive polynomial $q(n)$.*

*Proof.* We combine Lemma 5.1 with the proof method of lemma 5.2. First, we may define a function

$$a(x,y) \stackrel{\text{DEF}}{=} \begin{cases} 1 & \text{if } y = c_A(x) \\ 0 & \text{otherwise} \end{cases}$$

then we can amplify the accepting probabilities as in Lemma 5.1. We replicate the circuit enough times to get probabilities $\frac{8}{9}$ and $\frac{1}{9}$, so we have

For $x \in L$, we have $\sum_{y:a(x,y)=1} \left|\langle y|T|c_I(x)\rangle\right|^2 \geq \frac{8}{9}$.

For $x \notin L$, we have $\sum_{y:a(x,y)=1} = \left|\langle y|T|c_I(x)\rangle\right|^2 \leq \frac{1}{9}$.

Then we add an extra wire in state $|0\rangle$. To the end of the resulting circuit, we add a unit $f$ that will set the state $|1\rangle$ in the extra circuit iff more than half of the circuits output $|c_A(x)\rangle$, and otherwise leave it untouched. Then we will rollback the algorithm adding an inverse circuit of the original one to each subregister. The rollback is necessary because we want to have only one accepting configuration for each input $x$. The situation is illustrated in figure 5.1. Formally, we define

$$c'_I(x) \stackrel{\text{DEF}}{=} |c_I(x)^k, 0\rangle$$
$$c'_A(x) \stackrel{\text{DEF}}{=} |c_(x)^k, 1\rangle$$
$$T' \stackrel{\text{DEF}}{=} \left(\left(\bigotimes_k T\right) \otimes 1\right) f \left(\left(\bigotimes_k T^{-1}\right) \otimes 1\right)$$

and get

$$\langle c'_A(x)|T'|c'_I(x)\rangle = \langle c_I(x)^k, 1| \left(\left(\bigotimes_k T\right) \otimes 1\right) f \left(\left(\bigotimes_k T^{-1}\right) \otimes 1\right) |c_I(x)^k, 0\rangle =$$

$$\sum_{y:f(y)=1} \langle c'_I(x)^k, 1| \left(\left(\bigotimes_k T\right) \otimes 1\right) f|y, 0\rangle\langle y, 0| \left(\left(\bigotimes_k T^{-1}\right) \otimes 1\right) |c_I(x)^k, 0\rangle$$

$$\sum_{y:f(y)=1} \langle c'_I(x)^k, 1| \left(\left(\bigotimes_k T\right) \otimes 1\right) |y, 1\rangle\langle y, 0| \left(\left(\bigotimes_k T^{-1}\right) \otimes 1\right) |c_I(x)^k, 0\rangle$$

$$\sum_{y:f(y)=1} \langle c'_I(x)^k| \left(\bigotimes_k T\right) |y\rangle\langle y| \left(\bigotimes_k T^{-1}\right) |c_I(x)^k\rangle$$

$$\sum_{y:f(y)=1} \langle c'_I(x)^k| \left(\bigotimes_k T\right) |y\rangle\langle c_I(x)^k| \left(\bigotimes_k T^*\right) |y\rangle$$

$$\sum_{y:f(y)=1} \left|\langle c'_I(x)^k| \left(\bigotimes_k T\right) |y\rangle\right|^2$$

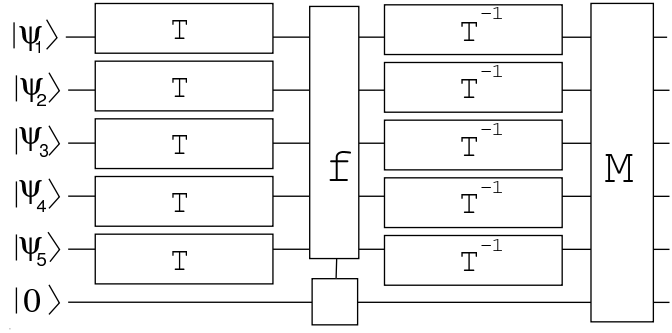

Figure 5.1: *The circuit used to amplify the probabilities in Lemma 5.3. Here we replicate the original circuit 5 times.*

Now we define new quantities $z_i$ the following way:

$$z_i = \begin{cases} 1 & \text{if } y_i = c_A(x) \\ 0 & \text{otherwise} \end{cases}$$

For $x \in L$, we have that $E(z) \geq 1 \cdot \frac{2}{3} + 0 \cdot \frac{1}{3} = \frac{2}{3}$. Thus,

$$\langle c'_A(x)|T'^{-1}fT'|c'_I(x)\rangle = \sum_{y:f(y)=1} \left|\langle y,0|T'|c_I(x)^k,0\rangle\right|^2 = P\left(\sum_{i=1}^k x_i \geq \frac{k}{2}\right)$$

$$\geq P\left(\left|\frac{\sum_{i=1}^k x_i}{k} - \frac{2}{3}\right| \leq \frac{1}{6}\right) = 1 - P\left(\left|\frac{\sum_{i=1}^k x_i}{k} - \frac{2}{3}\right| \geq \frac{1}{6}\right) \geq 1 - e^{-(\frac{1}{6})^2 \frac{k}{2}}$$

Similarly, for $x \notin L$, we have that $E(z) \leq 1 \cdot \frac{1}{3} + 0 \cdot \frac{2}{3} = \frac{1}{3}$. Thus,

$$\langle c'_A(x)|T'^{-1}fT'|c'_I(x)\rangle = \sum_{y:f(y)=1} \left|\langle y,0|T'|c_I(x)^k,0\rangle\right|^2 = P\left(\sum_{i=1}^k x_i \geq \frac{k}{2}\right)$$

$$\leq P\left(\left|\frac{\sum_{i=1}^k x_i}{k} - \frac{1}{3}\right| \geq \frac{1}{6}\right) \leq e^{-(\frac{1}{6})^2 \frac{k}{2}}$$

To finish the proof, it now suffices to set $k(n) = 36q(n)$. □

## 5.2 The new families

When we were treating the model of a quantum Turing machine in Chapter 2, we encountered transition matrices of the machines. They have a size exponential in the length of the input, but are unitary and each of their entries is just a simple function, computable in polynomial time. In Chapter 3, there was another family of unitary matrices, those that are expressible as a polynomial size circuits. Here, we will formalize these two families of unitary matrices and then use the formalism to define new complexity classes.

**Definition 5.2 (Unitary serie).** *An infinite series $U_1, U_2, \ldots$ of unitary matrices is called here a unitary series.*

Now, we will formally define two properties of unitary series. The first one, a property of being a polynomial size circuit, will be called $P1$.

**Definition 5.3 (P1).** *A unitary series $U_1, U_2, \ldots$ has property P1 iff there exist polynomials $p(i)$, $q(i)$ and a function $f(i)$ computable in time $q(i)$ such that for all $i$: $f(i) = U_i$ and $U_i$ is expressible as a circuit of size $p(i)$ of at most two-qubit gates.*

This property is very hard to manipulate with, since there are no general theorems available that would, seeing a huge matrix, tell us an effective way to find out whether it can be decomposed to some number of smaller gates or not. The only way is to try to decompose the matrix, which indeed takes time polynomial in dimension of that gate, which usually has exponential size itself.

When we defined a quantum Turing machine, we dealt with series of transition matrices, which were also unitary, but had another property, here called $P2$:

**Definition 5.4 (P2).** *A unitary series $U_1, U_2, \ldots$ has property P2 iff there exists a polynomial $p(i)$ and function $f(i, x, y)$ such that for all $i$:*

$$\forall x, y : \langle y|U_i|x\rangle = f(i, x, y)$$

*and the running time of $f$ is at most $p(i)$.*

We know from Section 2.1.1 that for a unitary series, having property $P2$ is not sufficient to be a series of transition matrices of some quantum Turing machines. On the other side, the property $P2$ is very easy to handle and if we just require this and not the existence of a quantum Turing machine which would correspond to that matrix, we arrive at another complexity class which we know is stronger than or equal to $BQP$ and thus we can eventually use this class to prove negative theorems about $BQP$.

### 5.2.1 The classes $BQ^{t(n)}$

**Definition 5.5** ($BQ^{t(n)}$). *Let us have an arbitrary function $t(n)$ on integers. A language $L$ is in $BQ^{t(n)}$ iff there is a unitary series $T_1, T_2, \ldots$ having property $P2$ and functions $c_I(x)$ and $c_A(x)$ computable in polynomial time such that for each $x$ and its length $n$:*

*For $x \in L$, we have $\left|\langle c_A(x)|T_n^{t(n)}|c_I(x)\rangle\right|^2 \geq \frac{2}{3}$.*

*For $x \notin L$, we have $\left|\langle c_A(x)|T_n^{t(n)}|c_I(x)\rangle\right|^2 \leq \frac{1}{3}$.*

Since we multiply a matrix with two vectors from each side, we silently assume their dimensions are compatible. If we used the property $P1$ instead of $P2$ and some polynomial $t(n)$, this definition would be equivalent to the definition of $BQP$ (see def. 5.1). The reason is that if we have polynomial size circuit, we can put a polynomial amount of its identical copies, one after another, and we still have a polynomial size circuit. Thus, any polynomial power in the definition would not change anything. So we can roughly say that we took the definition of $BQP$ and exchanged the two properties $P1$ and $P2$.

A circuit with property $P2$ is much different from a circuit with the property $P1$. We generally do not know how to efficiently implement a given circuit with property $P2$, which means finding a polynomial size circuit being equivalent to it. On the other side if we want to know a specific entry of the matrix representing that circuit, this can be answered in a polynomial time with respect to the length of the input. It is maybe a bit reminiscent of a black box approach: we do not know what a circuit composed of elementary gates would look like, but we can use a function to compute any entry of the building block.

In practice, there is often the situation that we do not know what particular output we are looking for, but have some property in mind we would like the outputs to have. This situation is expressed in the following definition. We have some function $a(x, y)$ which decides for a given $x$ at the input, if $y$ is one of the "good" outputs we are looking for or not. Then we want that for "good" $x$s, e.g. for the $x$s that are in the language we are interested in, that the probability that we obtain one of such be high, but not for the bad $x$s.

### 5.2.2 The classes $MQ^{t(n)}$

**Definition 5.6** ($MQ^{t(n)}$). *Let us have an arbitrary function $t(n)$ on integers. A language $L$ is in $MQ^{t(n)}$ iff there is a unitary series $T_1, T_2, \ldots$ having property $P2$ and functions $c_I(x)$ and $a(x, y)$ computable in polynomial time such that for each $x$ and its length $n$:*

*For $x \in L$, we have $\sum_{y:a(x,y)=1}\left|\langle y|T_n^{t(n)}|c_I(x)\rangle\right|^2 \geq \frac{2}{3}$.*

*For $x \notin L$, we have $\sum_{y:a(x,y)=1}\left|\langle y|T_n^{t(n)}|c_I(x)\rangle\right|^2 \leq \frac{1}{3}$.*

In the following text, we will omit the subscripts $n$ on the matrix when it will be clear about which $n$ we are talking.

In the class $BQP$, it does not matter whether we have one or more accepting configuration (see Lemma 5.1). Here, we can not use the same trick, since generally we can not put there an inverse of the matrix since we want that there is only one matrix repeating $t(n)$ times. Furthermore, having a function that computes the entries does not immediately imply we can a have a function computing the entries of the inverse. Therefore the hierarchies of $BQ^{t(n)}$ and $MQ^{t(n)}$ will generally be distinct.

Sometimes, we will let the classes $MQ$ and $BQ$ to access oracles and thus define a relativized versions of these classes.

**Definition 5.7** ($BQ$ **relativized**). *Let us have an arbitrary function $t(n)$ on integers. A language $L$ is in $BQ^{t(n)^O}$ for $O$ being an arbitrary oracle over the binary alphabet, iff there is a unitary series $T_1, T_2, \ldots$ having property $P2$ and functions $c_I(x)$ and $c_A(x)$ computable in polynomial time such that for each $x$ and its length $n$ the following holds:*

For $x \in L$, we have $\left|\langle c_A(x)|T_n^{t(n)}|c_I(x)\rangle\right|^2 \geq \frac{2}{3}$.

For $x \notin L$, we have $\left|\langle c_A(x)|T_n^{t(n)}|c_I(x)\rangle\right|^2 \leq \frac{1}{3}$.

*Here the entries of the matrices $T$ may be dependent on a polynomial amount of answers of the oracle $O$, but stay unitary regardless of the answers.*

**Definition 5.8 ($MQ$ relativized).** *Let us have an arbitrary function $t(n)$ on integers. A language $L$ is in $MQ^{t(n)^O}$ for $O$ being an arbitrary oracle over the binary alphabet, iff there is a unitary series $T_1, T_2, \ldots$ having property $P2$ and functions $c_I(x)$ and $a(x,y)$ computable in polynomial time for each $x$ and its length $n$:*

For $x \in L$, we have $\sum_{y:a(x,y)=1}\left|\langle y|T_n^{t(n)}|c_I(x)\rangle\right|^2 \geq \frac{2}{3}$.

For $x \notin L$, we have $\sum_{y:a(x,y)=1}\left|\langle y|T_n^{t(n)}|c_I(x)\rangle\right|^2 \leq \frac{1}{3}$.

*Here the entries of the matrices $T$ may be dependent on a polynomial amount of answers of the oracle $O$, but stay unitary regardless of the answers.*

## 5.3 Properties of the new classes

In this section, we will show some properties of the newly defined classes. In the first subsection, there will be self standing lemmas and observations meant to illustrate the meaning of the definitions and the properties $P1$ and $P2$. In the second subsection, we will state lemmas and theorems which will be used to build a hierarchy of the complexity classes.

### 5.3.1 Basic properties

**Observation 5.4.** *For all $t(n)$ it holds $BQ^{t(n)} \subseteq MQ^{t(n)}$.*

*Proof.* From the definitions. If we set $a(x,y) = 1 \iff c_A(x) = y$, we obtain a special case of $MQ^{t(n)}$ which equals $BQ^{t(n)}$. □

**Observation 5.5.** $BQ^1 = P$.

*Proof.* We want to show that $BQ^1 \subseteq P$. For each language $L \in BQ^1$ we have the following:

For $x \notin L$, we have $|\langle c_I(x)|T|c_A(x)\rangle|^2 \geq \frac{2}{3}$.

For $x \notin L$, we have $|\langle c_I(x)|T|c_A(x)\rangle|^2 \leq \frac{1}{3}$.

Here $T$ is a matrix having property $P2$ and $c_I$ and $c_A$ are functions in $P$. This implies that $f(x) := |\langle c_I(x)|T|c_A(x)\rangle|^2$ is also in $P$ and this implies that $L \in P$. □

**Observation 5.6.** $(P1 \to P2) \to (BQP \subseteq P)$.

*Proof.* Using $P1 \to P2$ we get any circuit having property $P1$, which is the case of $BQP$, can be realized by a matrix having property $P2$, thus $BQP \subseteq BQ^1 = P$. □

**Observation 5.7.** $(P2 \to P1) \to (BQ^{p(n)} \subseteq BQP)$.

*Proof.* Immediate, see Definition 5.5. □

**Observation 5.8.** $BQP \subseteq BQ^{poly(n)}$

*Proof.* Each problem in $BQP$ has a quantum Turing machine solving it. And each quantum Turing machine has a series of transition matrices that has property $P2$ and runs in a polynomial time. □

As we were able to amplify the accepting probabilities in the case of $BQP$ (Lemma 5.3), we would like to be able to amplify the probabilities in case of $BQ^{t(n)}$ and $MQ^{t(n)}$. This would justify their definitions of them and show their robustness. However, the tricks we could do with matrices with property $P1$ can not be straightforwardly applied to matrices with property $P2$. For example, generally we can not compute an inverse of an exponentially large matrix in polynomial time. Therefore the question of whether it is possible to amplify the class $BQ^{t(n)}$ (for nontrivial cases $t(n) \neq 1$) is left open here. On the other side, the class $MQ^{t(n)}$ can be easily amplified because it allows for multiple configuration so we can use exactly the same trick as in Lemma 5.1.

**Lemma 5.9 (Amplification of $MQ$).** *The probabilities $\frac{2}{3}$ and $\frac{1}{3}$ in the definition of $MQ^{t(n)}$ may be equivalently changed to any $a$ and $b$ such that $a \geq b + \epsilon$ for some $\epsilon > 0$.*

*Proof.* Let $L$ be a language in $MQ^{t(n)}$ with the definition changed to $a$ and $b$ satisfying the assumption. Let $M$ be the matrix from definition 5.6 showing that $L \in MQ^{t(n)}$ with the altered probabilities. We will chose an even polynomial $k(n)$ and make a tensor product of $k(n)$ copies of matrices $M$. As in Lemma 5.3, we will accept iff at least $\frac{a+b}{2}$ of the copies accepts. The new function $a'(x, y_1 y_2 \ldots y_{k(n)})$ will check that for at least $\frac{a+b}{2}$ of the $y$s there is $a(x, y) = 1$. We have

$$T_k \equiv \bigotimes_{i=1}^{k} M$$

and

$$a'(x, y_1 y_2 \ldots y_{k(n)}) \stackrel{\text{DEF}}{=} \begin{cases} 1 & \text{if for at least } \frac{a+b}{2} \text{ of the } y\text{s there is } a(x,y) = 1 \\ 0 & \text{otherwise} \end{cases}$$

We define $z_i$ as in Lemma 5.3 and proceed analogously. Then we have

for $x \in L$:

$$\sum_{y: a'(x, y_1 y_2 \ldots y_{k(n)}) = 1} \left| \langle y | T^{t(n)} | c_I(x) \rangle \right|^2 = P\left( \sum_{i=1}^{k} z_i \geq \frac{k(n)}{2} \right) \geq 1 - e^{-(\frac{a+b}{2})^2 \frac{k}{2}}$$

for $x \notin L$:

$$\sum_{y: a'(x, y_1 y_2 \ldots y_{k(n)}) = 1} \left| \langle y | T^{t(n)} | c_I(x) \rangle \right|^2 = P\left( \sum_{i=1}^{k} z_i \leq \frac{k(n)}{2} \right) \leq e^{-(\frac{a+b}{2})^2 \frac{k}{2}}$$

To finish the proof, it suffices to notice that the gap between the completeness and soundness gets exponentially bigger with increasing $k(n)$, so we can reach any other desired $a'$ and $b'$ such that $a' \geq b' + \epsilon$ for some $\epsilon > 0$. □

### 5.3.2 Building the hierarchy

In this subsection, we will state theorems and lemmas needed to build the hierarchy of the classes $BQ^{f(n)}$ and $MQ^{f(n)}$.

Except for the class $BQ^1$, which was shown to be equal to $P$ in observation 5.3.1, $MQ^1$ will be the lowest class in the hierarchy, because we will show later on that $MQ^1 \subseteq BQ^2$. Still, it is already superior to $BPP$, the class that is believed to encompass all efficiently solvable problems.

**Observation 5.10.** $MQ^1 \supseteq BPP$.

*Proof.* Let us have a language $L$ in $BPP$, a polynomial $p(n)$, and a probabilistic Turing machine $M$ with running time $p(n)$ accepting $L$, proving that $L \in BPP$. Let $r(n)$ be the amount of random bits used for inputs of length $n$. We know that $r(n) \leq p(n)$. Let $M(x, r)$ denote the output (either 1 or 0) of $M$ over input $x$ and random bits $r$. Here 1 stays for accepting configuration and 0 for rejecting. Then the probability of acceptance over $x$ reads:

$$P(x) = \frac{1}{2^{r(|x|)}} \sum_{y:|y|=r(|x|)} M(x, y)$$

We can straightforwardly realize this probability of acceptance by running the machine $M$ on all possible $y$'s of length $r(|x|)$ with equal amplitude, and then sum up the results with help of accepting function $a(x, (y, z))$, that will accept iff $M(x, z)$ will accept. For obtaining equal amplitudes, we will use the Hadamard gate on each qubit.

The probability of acceptance, $P(x)$, will be thus obtained defining

$$c_I(x) \stackrel{\text{DEF}}{=} |x, 0^{r(|x|)}\rangle,$$
$$a(x, (y, z)) \stackrel{\text{DEF}}{=} M(x, z),$$
$$\langle i|T|j\rangle \stackrel{\text{DEF}}{=} \frac{1}{\sqrt{2^n 2^{r(n)}}} (-1)^{\sum_k i_k j_k},$$

where $i_k$ is the $k$-th bit of $i$, because we have that

$$\sum_{(y,z):a(x,(y,z))=1} \left|\langle y, z|T|c_I(x)\rangle\right|^2 = \sum_{(y,z):M(x,z)=1} \left|\langle y, z|T|x, 0^{r(|x|)}\rangle\right|^2 =$$

$$\sum_y \sum_{z:M(x,z)=1} \left|\frac{1}{\sqrt{2^{|x|} 2^{r(|x|)}}} (-1)^{\sum_i y_i x_i}\right|^2 =$$

$$\sum_y \sum_{z:M(x,z)=1} \frac{1}{2^{|x|} 2^{r(|x|)}} = \sum_{z:M(x,z)=1} \frac{1}{2^{r(|x|)}} = P(x)$$

$\square$

Later on, with the help of $GapP$ functions, we will state stronger theorems about the classes $MQ^{p(n)}$ for a polynomial $p(n)$, which will immediately lead to the following observation as a corollary. We still mention this observation here because it provides a good insight into the classes.

**Observation 5.11.** *For each polynomial $p(n)$, the class $MQ^{p(n)} \subseteq PSPACE$.*

*Proof.* It suffices to show that for any functions $c_I(x)$ and $a(x, y)$ computable in polytime and a matrix series $T$ having property $P2$, the logical function

$$F(x, n) \equiv \left(\sum_{y:a(x,y)=1} |\langle y|T^{p(n)}|c_I(x)\rangle|^2 \geq \frac{2}{3}\right)$$

defines a language $L = \{(x, n)|F(x, n) = 1\}$ that is in $PSPACE$. We have

$$\langle y|T^{p(n)}|c_I(x)\rangle = \sum_{i_1, i_2, \ldots i_{p(n)}} \langle y|T|i_1\rangle\langle i_1|T|i_2\rangle \ldots \langle i_{p(n)-1}|T|i_{p(n)}\rangle\langle i_{p(n)}|T|c_I(x)\rangle$$

Each element of the sum is a function computable in polytime with respect to $n$, thus computing it can not require more than a polynomial amount of space. What we need to remember when computing the sum is: The actual sum and indices $i_1, i_2, \ldots i_{p(n)}$. This requires only size polynomial with respect to $n$. To compute the sum $\sum_{y:a(x,y)=1} |\langle y|T^{p(n)}|c_I(x)\rangle|^2$, it suffices to remember the actual index $y$. $\square$

The following lemma is a handy tool telling us that we do not have to bother with hermitian conjugates of the matrices in the definition of the class $MQ^1$, since all of them can be defined as hermitian.

**Lemma 5.12.** *Without loss of generality, me may require that the matrix $T$ in the definition of $BQ^1$ and $MQ^1$ be hermitian.*

*Proof.* Let us have a matrix $T$ and functions $c_I(x)$ and $c_A(x)$. We then define a new matrix $S$ as

$$\langle y, 0|S|x, 0\rangle \stackrel{\text{DEF}}{=} \langle y, 1|S|x, 1\rangle = 0$$
$$\langle y, 0|S|x, 1\rangle \stackrel{\text{DEF}}{=} \langle y|T|x\rangle$$
$$\langle y, 1|S|x, 0\rangle \stackrel{\text{DEF}}{=} \langle y|T^\dagger|x\rangle$$

The new matrix is illustrated in Figure 5.2 a). Such a matrix is clearly hermitian. It is also unitary, since we have

$$S = \begin{pmatrix} 0 & T \\ T^\dagger & 0 \end{pmatrix}$$
$$SS^\dagger = S^2 = \begin{pmatrix} TT^\dagger & 0 \\ 0 & T^\dagger T \end{pmatrix} = \begin{pmatrix} 1 & 0 \\ 0 & 1 \end{pmatrix}$$

Now, we are ready to define

$$c'_I(x) \stackrel{\text{DEF}}{=} |c_I(x), 1\rangle$$
$$c'_A(x) \stackrel{\text{DEF}}{=} |c_A(x), 0\rangle$$
$$a'(x, (y, i)) \stackrel{\text{DEF}}{=} a(x, y)\delta_{i0}$$

Then we have $\langle c'_A(x)|S|c'_I(x)\rangle = \langle c_A(x)|T|c_I(x)\rangle$ and thus

$$|\langle c'_A(x)|S|c'_I(x)\rangle|^2 = |\langle c_A(x)|T|c_I(x)\rangle|^2$$

and

$$\sum_{(y,i):a'(x,(y,i))=1} |\langle y, i|S|c'_I(x)\rangle|^2 = \sum_{y:a(x,y)=0} |\langle y|T|c_I(x)\rangle|^2$$

□

In the definition of the classes $BQ^{t(n)}$ and $MQ^{t(n)}$, the function $t(n)$ reflects the power of the class and complexity of the problems in it. The higher the value $t(n)$, the more work we have to do to obtain the result. One would thus expect that for each integer $i$, $BQ^i \subseteq BQ^{i+1}$ and $MQ^i \subseteq MQ^{i+1}$. However, it is not obvious why this should hold. If we for example imagine a language $L$ in $MQ^7$, it is not clear how to accept it in $MQ^8$, because if we take a matrix $M$ that serves to prove $L \in MQ^7$, once we compute its 8th power, its entries will typically be very different and useless. What would save us is if we could construct another matrix from the matrix $M$ such that in its 8th power, there are somewhere the entries of the 7th power of $M$, or their simple transformations, like multiples etc. Here in the two following theorems we construct such composite matrices to show $MQ^1 \subseteq MQ^k$ for any $k \geq 1$. Other such relations, for example whether also $MQ^2 \subseteq MQ^k$ for any $k \geq 2$ are left open here.

**Theorem 5.13.** *The following holds:*

$$MQ^1 \subseteq MQ^{1+4k}$$
$$MQ^1 \subseteq MQ^{2+4k}$$
$$MQ^1 \subseteq MQ^{3+4k}$$

*for any nonnegative integer $k$.*

Figure 5.2: The new matrices for the proofs of lemmas 5.12 and 5.13

*Proof.* Let us have a language $L \in MQ^1$, a matrix $T$ and functions $c_I(x)$ and $a(x,y)$ solving it. We know from Lemma 5.12 that we may assume the matrix $T$ is hermitian. That means we have $T = T^\dagger$ and $TT^\dagger = T^2 = 1$. Then, let us define a matrix $S$ as

$$\langle y, 0|S|x, 0\rangle \stackrel{\text{DEF}}{=} \langle y, 1|S|x, 1\rangle = \frac{1}{\sqrt{2}}\langle y|T|x\rangle$$

$$\langle y, 0|S|x, 1\rangle \stackrel{\text{DEF}}{=} \frac{1}{\sqrt{2}}\delta_{xy}$$

$$\langle y, 1|S|x, 0\rangle \stackrel{\text{DEF}}{=} -\frac{1}{\sqrt{2}}\delta_{xy}$$

The new matrix is illustrated in Figure 5.2 b). Such a matrix is unitary, because we have

$$S = \frac{1}{\sqrt{2}}\begin{pmatrix} T & 1 \\ -1 & T \end{pmatrix}$$

$$S^\dagger = \frac{1}{\sqrt{2}}\begin{pmatrix} T^\dagger & -1 \\ 1 & T^\dagger \end{pmatrix} = \frac{1}{\sqrt{2}}\begin{pmatrix} T & -1 \\ 1 & T \end{pmatrix} = S^T$$

$$SS^\dagger = \frac{1}{2}\begin{pmatrix} TT^\dagger + 1 & -T + T^\dagger \\ -T^\dagger + T & (-1)(-1) + TT^\dagger \end{pmatrix} = \frac{1}{2}\begin{pmatrix} 2 & 0 \\ 0 & 2 \end{pmatrix} = 1$$

Now, we are ready to define $c'_I(x) \stackrel{\text{DEF}}{=} |c_I(x), 1\rangle$ and $a'(x, (y, i)) \stackrel{\text{DEF}}{=} \delta_{i0}a(x, y)$. Then, as

$$S^2 = \frac{1}{2}\begin{pmatrix} T^2 - 1 & T + T \\ -T - T & -1 + T^2 \end{pmatrix} = \begin{pmatrix} 0 & T \\ -T & 0 \end{pmatrix}$$

we get

$$\sum_{(y,i):a'(x,(y,i))=1} \left|\langle y, i|S^2|c_I(x), 1\rangle\right|^2 = \sum_{y:a(x,y)=1} \left|\langle y|T|c_I(x)\rangle\right|^2$$

showing $MQ^1 \subseteq MQ^2$.

Alternatively, when we define $c'_I(x) \stackrel{\text{DEF}}{=} |c_I(x), 0\rangle$ and $a'(x, (y, i)) \stackrel{\text{DEF}}{=} \delta_{i0}a(x, y)$, then since

$$S^3 = \frac{1}{\sqrt{2}}\begin{pmatrix} T & 1 \\ -1 & T \end{pmatrix}\begin{pmatrix} 0 & T \\ -T & 0 \end{pmatrix} = \frac{1}{\sqrt{2}}\begin{pmatrix} -T & 1 \\ -1 & -T \end{pmatrix}$$

we get

$$\sum_{(y,i):a'(x,(y,i))=1} \left|\langle y, i|S^2|c_I(x), 1\rangle\right|^2 = \frac{1}{2}\sum_{y:a(x,y)=1} \left|\langle y|T|c_I(x)\rangle\right|^2$$

So, instead of accepting probabilities $\frac{2}{3}$ and $\frac{1}{3}$, we would obtain $\frac{1}{3}$ and $\frac{1}{6}$ respectively. However, this can be cured by lemma 5.9 to finally show that $MQ^1 \subseteq MQ^3$.

It also holds that $S^4 = -1$, since we have that

$$S^4 = \begin{pmatrix} 0 & T \\ -T & 0 \end{pmatrix}^2 = \begin{pmatrix} -1 & 0 \\ 0 & -1 \end{pmatrix}$$

Thus, adding a power of 4 will only change the global phase of the resulting matrix. That means we may add a power of any multiple of 4 to the equations 5.3.2 and 5.3.2 without changing the right hand side. This proves the rest. □

**Theorem 5.14.** *The following holds for any nonnegative integer $k$:*

$$MQ^1 \subseteq MQ^{4k}$$

*Proof.* We will use exactly the same technique as in the theorem 5.13, only the matrix will be a little bit more complicated. Let us have a language $L \in MQ^1$, a matrix $T$ and functions $c_I(x)$ and $a(x,y)$ solving it. We know from Lemma 5.12 that we may assume that the matrix $T$ is hermitian. We define a matrix $R$ as

$$R = \frac{1}{\sqrt{2}} \begin{pmatrix} S & H \\ H & -S \end{pmatrix}$$

where $S$ is the matrix from the proof of Lemma 5.13 and $H$ is the Hadamard matrix. The matrix $R$ is unitary, because we have

$$R^\dagger = \frac{1}{\sqrt{2}} \begin{pmatrix} S^T & H \\ H & -S^T \end{pmatrix}$$

$$RR^\dagger = \frac{1}{2} \begin{pmatrix} 2 & SH - HS^T \\ -SH + HS^T & 2 \end{pmatrix}$$

and

$$HS^T = \frac{1}{2} \begin{pmatrix} 1 & 1 \\ 1 & -1 \end{pmatrix} \begin{pmatrix} T & -1 \\ 1 & T \end{pmatrix}$$
$$= \begin{pmatrix} T+1 & T-1 \\ T-1 & -1-T \end{pmatrix}$$
$$= \frac{1}{2} \begin{pmatrix} T & 1 \\ -1 & T \end{pmatrix} \begin{pmatrix} 1 & 1 \\ 1 & -1 \end{pmatrix} = SH$$

It can be verified that

$$R^4 = \frac{1}{2} \begin{pmatrix} -1 & T & 1 & -1 \\ -T & -1 & -1 & -1 \\ -1 & 1 & -1 & T \\ 1 & 1 & -T & -1 \end{pmatrix}$$

Now, we are ready to define $c'_I(x) \stackrel{\text{DEF}}{=} |c_I(x), 1\rangle$ and $a'(x, (y, i)) \stackrel{\text{DEF}}{=} \delta_{i0} a(x, y)$, so that means we will look onto the second column and the first row. We get

$$\sum_{(y,i): a'(x,(y,i))=1} \left| \langle y, i | R^4 | c_I(x), 1 \rangle \right|^2 = \frac{1}{4} \sum_{y: a(x,y)=1} \left| \langle y | T | c_I(x) \rangle \right|^2.$$

the probabilities $\frac{2}{3}$ and $\frac{1}{3}$ are changed by a factor of $\frac{1}{4}$, but we may still get them back using Lemma 5.9. □

**Corollary 5.15.** *For any integer $k \geq 1$, $MQ^1 \subseteq MQ^k$.*

*Proof.* Follows from Theorems 5.13 and 5.14. □

In the case of $BQP$, we could equivalently allow for multiple accepting configurations. On the contrary, here its not clear whether $BQ^{t(n)} = MQ^{t(n)}$ for some $t(n)$, not even whether $MQ^{t(n)} \subseteq BQ^{t'(n)}$ for some $t'(n)$. The only such question answered here is the following:

**Theorem 5.16.** $MQ^1 \subseteq BQ^2$.

*Proof.* Let $T$, $c_I$ and $a(x,y)$ be the items as in the definition 5.6 that give the evidence for $L \in MQ^1$. We will use the amplification in Lemma 5.9 and assume we have

For $x \in L$: $\sum_{y:a(x,y)=1} |\langle y|T^{p(n)}|c_I(x)\rangle|^2 \geq \frac{8}{9}$.

For $x \notin L$: $\sum_{y:a(x,y)=1} |\langle y|T^{p(n)}|c_I(x)\rangle|^2 \leq \frac{1}{9}$.

Furthermore, we will assume that for each $i, j$, we have

$$\langle i|T|j\rangle = \langle j|T^*|i\rangle.$$

Such an assumption is justified by lemma 5.12, because we may assume that the matrix $T$ is hermitian and have

$$\langle i|T|j\rangle = \langle j|T^T|i\rangle = \langle j|(T^\dagger)^*|i\rangle = \langle j|T^*|i\rangle.$$

We will define a matrix $M$ and functions $c'_I(x)$, $c'_A(x)$ as

$$\langle i,b|M|j,c\rangle \stackrel{\text{DEF}}{=} \frac{1}{\sqrt{2}}(-1)^{c_i b_i}(-1)^{b(1-a(x,i))}\langle i|T|j\rangle$$

$$|c'_I(x)\rangle \stackrel{\text{DEF}}{=} |c_I(x), 0\rangle$$

$$|c'_A(x)\rangle \stackrel{\text{DEF}}{=} |c_I(x), 0\rangle$$

Such a matrix is unitary, because it is a tensor product of the Hadamard matrix and the matrix $T$ (which is unitary by assumption), with some rows multiplied by $-1$. Then we have

$$\langle c'_A(x)|M^2|c'_I(x)\rangle = \sum_{i,b}\langle c_I(x),0|M|i,b\rangle\langle i,b|M|c_I(x),0\rangle =$$

$$\sum_{i,b}\frac{1}{\sqrt{2}}(-1)^{0_i,b_i}(-1)^{0(1-a(x,c_I(x)))}\langle c_I(x)|T|i\rangle\frac{1}{\sqrt{2}}(-1)^{0_i,b_i}(-1)^{b(1-a(x,i))}\langle i|T|c_I(x)\rangle =$$

$$\frac{1}{2}\sum_{i,b}(-1)^{b(1-a(x,i))}\left|\langle i|T|c_I(x)\rangle\right|^2 =$$

$$\frac{1}{2}\Big(\sum_i(-1)^{0(1-a(x,i))}\left|\langle i|T|c_I(x)\rangle\right|^2 + \sum_i(-1)^{1(1-a(x,i))}\left|\langle i|T|c_I(x)\rangle\right|^2\Big) =$$

$$\frac{1}{2}\Big(1 + \sum_{i:a(x,i)=1}\left|\langle i|T|c_I(x)\rangle\right|^2 - \sum_{i:a(x,i)=0}\left|\langle i|T|c_I(x)\rangle\right|^2\Big) =$$

$$\frac{1}{2}\Big(2\sum_{i:a(x,i)=1}\left|\langle i|T|c_I(x)\rangle\right|^2\Big) = \sum_{i:a(x,i)=1}\left|\langle i|T|c_I(x)\rangle\right|^2$$

Then for $x \in L$, we have that

$$\left|\langle c'_A(x)|M|c'_I(x)\rangle\right|^2 = \Big|\sum_{y:a(x,y)=1}|\langle y|T|x\rangle|^2\Big|^2 \geq \left(\frac{8}{9}\right)^2 \geq \frac{2}{3}$$

and for $x \notin L$ we have that

$$\left|\langle c'_A(x)|M|c'_I(x)\rangle\right|^2 = \left|\sum_{y:a(x,y)=1}|\langle y|T|x\rangle|^2\right|^2 \leq \left(\frac{1}{9}\right)^2 \leq \frac{1}{3}$$

□

## 5.4 Putting some problems into the classes

Now we will accommodate some well known problems in the new classes, to illustrate their power. Since we will also ask what functions the classes can compute, we will first define what we mean by that:

**Definition 5.9.** *A function $f(x)$ is computed by a class $C$ iff for each index $i$ we have $L \stackrel{DEF}{=} \{<x,i>|f(x)_i = 1\} \in C$ where $f(x)_i$ denotes the $i$−th bit of $f(x)$.* [1]

**Definition 5.10 (NP-Search problems).** *A NP-Search problem is the following: For a given function $f(x,y)$ with the range of $\{0,1\}$ computable in time $p(n) = poly(n)$ and for a given $x$, output $g(x)$ such that*

$$\bigl(g(x) = y \ \& \ f(x,y) = 1\bigr) \lor \bigl(q(x) = 0 \ \& \ \forall y = \{0,1\}^{p(x)} : f(x,y) = 0\bigr)$$

As was shown in [4], the class $BQP$ can only lead to a polynomial speedup over the class $BPP$ when solving $NP$-Search problems. The following theorem shows that the classes $MQ^{p(n)}$ for some polynomial $p$ and even their union $MQ^{poly(n)}$ are no better in this aspect.

**Theorem 5.17.** $MQ^{poly(n)}$ *does not solve $NP$-Search problems given as oracle. More precisely:*
*For each $NP$-search problem $f(x,y)$ computable in a polynomial time, let us define an oracle*

$$O_f \stackrel{DEF}{=} \{(x,y)|f(x,y) = 1\}$$

*Then for each polynomial $p(n)$ there exists an $O_f$ such that*

$$L_i = \{<x,i>|g(x)_i = 1\} \notin MQ^{p(n)^{O_f}}$$

*Proof.* We will proceed using the polynomial method in a similar way as it was applied in [4]. From the definition of $MQ^{p(n)}$, we know that there exists a polynomial $r(n)$, such that $\langle y|T|x\rangle$ is a function of $x$ and $y$ computable in time $r(n)$. Thus, it can ask the oracle at most $r(n)$ times. As in [4], we may express each element $\langle y|T|x\rangle$ as a polynomial of variables indicating presence of pairs in the oracle of degree at most $r(n)$. It follows that $\langle y|T^{p(n)}|x\rangle$ is for each $x$ and $y$ a polynomial of degree at most $p(n)r(n)$ and its square of degree at most $2(p(n)r(n))$. It follows that

$$\sum_{y:a(x,y)=1}\left|\langle y|T^{p(n)}|c_I(x)\rangle\right|^2$$

is expressible as a polynomial of degree at most $2(p(n)r(n))$. However, the *or* function of $n$ variables is a polynomial of degree $n$. Since we have exponential amount of pairs, we can not realize their *or* function with a polynomial of a degree $2p(n)r(n)$, which is still only a polynomial. □

**Corollary 5.18.** *There exists oracle $O$ such that $BQ^{2^O} \not\supseteq NP^O$.*

---

[1] As with conventional quantum circuits, we will not prove directly that the language $L = \{<x,i> |f(x)_i = 1\}$ is in a class $C$, but only that the language $L_i = \{x|f(x)_i = 1\}$ is in $C$ for all $i$ and furthermore the algorithm that decides a language $L_i$ is for all $i$ the same and only takes $i$ as parameter.

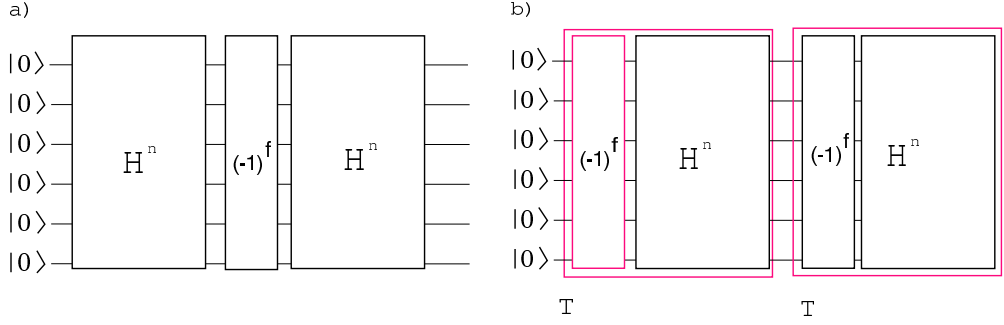

Figure 5.3: The trick used to fit the Deutsch-Jozsa algorithm into the $BQ^2$ class. In a), there is the original setup, in b) there is the one we use, which gives the same result up to a global phase.

Except for the trivial class $BQ^1$, which equals to $P$ (Observation 5.3.1), the lowest class in the hierarchy is the class $MQ^1$, because we have $MQ^1 \subseteq BQ^2$ in Theorem 5.16. The second lowest class is then obviously $BQ^2$. Yet it captures the famous Deutsch-Jozsa problem. For this problem, see [6].

**Theorem 5.19.** *The class $BQ^2$ solves the Deutsch-Jozsa problem. More precisely, given a function $f(x) \to \{0, 1\}$ that is either constant or balanced [2] and an oracle*

$$O_f = \{x | f(x) = 1\}$$

*the language*

$$L = \begin{cases} \{0^n | n \text{ arbitrary integer }\} & \text{if the function } f \text{ is constant} \\ \{\} & \text{if the function } f \text{ is balanced} \end{cases}$$

*is in $BQ^{2Q_f}$.*

*Proof.* We will realize the circuit in the Deutsch-Jozsza algorithm (see Figure 5.3 a)) by two equal matrices $T$ having the property $P2$ (see Figure 5.3 b)). The matrix $T$ will be a product of a matrix realizing $x \to (-1)^{f(x)}$, as in the original version of the algorithm, and $H^n$. We may add another matrix for the $f$ function to the front, since it will only add the number $(-1)^{f(0)}$ to the global phase and thus will not change the result. Formally, we define a matrix $T$ as

$$\langle y|T|x\rangle \stackrel{\text{DEF}}{=} (-1)^{f(x)}\langle y|H^n|x\rangle = \frac{1}{\sqrt{2^n}}(-1)^{f(x)}(-1)^{\sum_i x_i y_i \bmod 2}$$

which is obviously computable in polytime and unitary. We define $c_i(x) = 0^n$ and $c_A(x) = 0^n$ for $x$ of length $n$. Then we have

$$\left|\langle 0^n|T^2|0^n\rangle\right|^2 = \left|\sum_k \langle 0^n|T|k\rangle\langle k|T|0^n\rangle\right|^2 =$$

$$= \left|\sum_k (-1)^{f(k)}\frac{1}{\sqrt{2^n}}(-1)^{\sum_i k_i 0^n_i \bmod 2}\frac{1}{\sqrt{2^n}}(-1)^{f(0^n)}(-1)^{\sum_i 0^n_i k_i \bmod 2}\right|^2 =$$

$$= \left|\sum_k \frac{1}{2^n}(-1)^{f(k)}(-1)^{2\sum_i k_i 0^n_i \bmod 2}\right|^2 = \frac{1}{2^{2n}}\left|\sum_k (-1)^{f(k)}\right|^2$$

---

[2] With *balanced* function we mean a function that outputs 1 on exactly half of the inputs and zero on the other half.

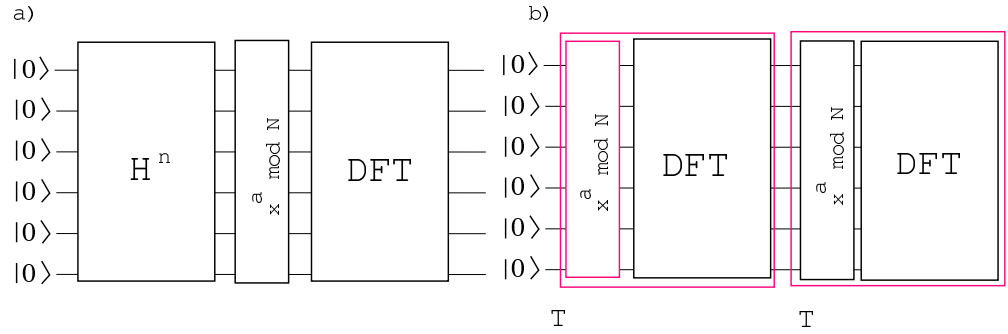

Figure 5.4: The trick used to fit the Shor algorithm into the $MQ^2$ class. In a), there is the original setup, in b) there is the one we use, which gives the same result up to a global phase.

If the function is constant, then the sum $\sum_k (-1)^{f(k)}$ equals $\pm 2^n$ and the probability $\left|\langle 0^n|T^2|0^n\rangle\right|^2$ equals 1. If the function is balanced, both the sum and the probability is 0. $\square$

**Corollary 5.20.** *There exists an oracle $O$ such that $BQ^{2^O} \nsubseteq NP$.*

Perhaps not surprisingly, to realize the famous Shor algorithm [14], we also need only to square a matrix, this time however we need to have multiple accepting configurations.

**Theorem 5.21.** *The class $MQ^2$ solves the factoring problem. More precisely, there exists a constant $k$ such that given numbers $x$ and $N > k$ as in the Shor algorithm, the language*

$$L = \{<N, i> | x^a \mod N \text{ has a period } r \text{ whose i-th bit is 1}\}$$

*is in $MQ^2$.*

*Proof.* We will use the same notation as in the original paper by Shor [14]. $N$ is the number to factorize and $N = p_1 p_2$ where both $p_1$ $p_2$ are primes and are different from each other. We then arbitrarily choose an $x$ coprime to $N$. The pair $(x, N)$ is the input of the algorithm. The goal is to find the smallest $r \neq 0$ such that $x^r \mod N = 1$. This $r$ is called a *period* of $x$. We choose a number $q$ such that $q$ is a power of 2 and $q \geq 2^{2\lceil \log_2 N \rceil}$. This number will, together with the length of $x$ and $N$, determine the size of the matrix.

As in the previous lemma, we will convert the circuit into the desired form by adding another unit to the front which will only affect the global phase. See Figure 5.4.

The space we will work with is spanned by vectors of the form $|x, N, a, i\rangle$, where the third and the fourth part of the vector are arbitrary integers from the interval $[0, q-1]$. From now on, if we do not specify the sums bounds, we mean summing over the range of the variable. We first define two matrices called $DFT$ and $MOD$ in the following manner:

$$\langle x', N', a', i'|DFT|x, N, a, i\rangle \stackrel{\text{DEF}}{=} \frac{1}{\sqrt{q}} \delta_{x,x'} \delta_{N,N'} \delta_{i,i'} e^{\frac{2i\pi}{q} aa'}$$

$$\langle x', N', a', i'|MOD|x, N, a, i\rangle \stackrel{\text{DEF}}{=} \frac{1}{\sqrt{q}} \delta_{x,x'} \delta_{N,N'} \delta_{a,a'} \delta_{i'+i, x^a \mod N}$$

Now, we may simply check that both the matrices are unitary:

$$\sum_{x',N',a',i'} \langle x_1, N_1, a_1, i_1|DFT|x', N', a', i'\rangle \langle x_2, N_2, a_2, i_2|\overline{DFT}|x', N', a', i'\rangle =$$

$$\sum_{x',N',a',i'} \frac{1}{q}\delta_{x_1,x'}\delta_{N_1,N'}\delta_{i_1,i'}e^{\frac{2i\pi}{q}a_1 a'}\delta_{x_2,x'}\delta_{N_2,N'}\delta_{i_2,i'}e^{-\frac{2i\pi}{q}a_2 a'} =$$

$$\sum_{a'} \frac{1}{q}\delta x_1, x_2 \delta_{N_1,N_2}\delta_{i_1,i_2}e^{\frac{2i\pi}{q}(a_1-a_2)a'} \delta x_1, x_2 \delta_{N_1,N_2}\delta_{i_1,i_2}\sum_{a'} e^{\frac{2i\pi}{q}(a_1-a_2)a'} =$$

$$\begin{cases} \frac{1}{q}\delta x_1, x_2 \delta_{N_1,N_2}\delta_{i_1,i_2}\frac{e^{\frac{2i\pi}{q}(a_1-a_2)q}-1}{e^{\frac{2i\pi}{q}(a_1-a_2)}-1} = 0 & \text{iff } a_1 \neq a_2 \\ \frac{1}{q}q = 1 & \text{otherwise} \end{cases}$$

$$= \delta x_1, x_2 \delta_{N_1,N_2}\delta_{i_1,i_2}\delta_{a_1,a_2}$$

$$\sum_{x',N',a',i'} \langle x_1, N_1, a_1, i_1|MOD|x', N', a', i'\rangle \langle x_2, N_2, a_2, i_2|MOD|x', N', a', i'\rangle =$$

$$\sum_{x',N',a',i'} \frac{1}{q}\delta_{x_1,x'}\delta_{N_1,N'}\delta_{a_1,a'}\delta_{i_1+i',x^{a'} \bmod N}\delta_{x_2,x'}\delta_{N_2,N'}\delta_{a_2,a'}\delta_{i'-i+2,x^{a'} \bmod N} =$$

$$\sum_{i'} \frac{1}{q}\delta_{x_1,x_2}\delta_{N_1,N_2}\delta_{a_1,a_2}\delta_{i_1+i',x^{a_1} \bmod N}\delta_{i'+i_2,x^{a_1} \bmod N} =$$

$$\sum_{i'} \frac{1}{q}\delta_{x_1,x_2}\delta_{N_1,N_2}\delta_{a_1,a_2}\delta_{i_2,i_1}$$

Now we are ready to define a unitary matrix $T \stackrel{\text{DEF}}{=} DFT \cdot MOD$. Its elements read

$$\langle x', N', a', i'|T|x, N, a, i\rangle = \delta_{x,x'}\delta_{N,N'}e^{\frac{2i\pi}{q}a'a}\delta_{i+i',x^a \bmod N}$$

and are clearly functions computable in polytime.

The function $c_I$ will be defined as

$$c_I(x, N) \stackrel{\text{DEF}}{=} |x, N, 0, 0\rangle$$

Now, we will compute the probability of outputting configuration $|x, N, n\frac{q}{r}, i\rangle$ for any integer $0 \leq n \leq r-1$ and any integer $0 \leq i \leq q-1$. We have

$$\sum_{(x',N',a',i'):a'=nq/r} \left|\langle x',N',a',i'|T^2|c_I(x,N)\rangle\right|^2 =$$

$$\sum_{\substack{(x',N',a',i'):\\a'=nq/r}} \left|\sum_{a'',i''} \langle x',N',a',i'|T|x',N',a'',i''\rangle\langle x',N',a'',i''|T|x,N,0,0\rangle\right|^2 =$$

$$\sum_{(a',i'):a'=nq/r} \left|\sum_{a'',i''} \langle x,N,a',i'|T|x,N,a'',i''\rangle\langle x,N,a'',i''|T|x,N,0,0\rangle\right|^2 =$$

$$\sum_{(a',i'):a'=nq/r} \left|\frac{1}{q}\sum_{a'',i''} e^{\frac{2i\pi}{q}a'a''}\delta_{i'+i'',x^{a''} \bmod N}e^0\delta_{i'',1}\right|^2 =$$

$$\sum_{(a',i'):a'=nq/r} \frac{1}{q^2}\left|\sum_{a'':x^{a''} \bmod N=i'+1} e^{\frac{2i\pi}{q}a'a''}\right|^2 =$$

$$\sum_{i'=0}^{r-1}\sum_{a':a'=nq/r} \frac{1}{q^2}\left|\sum_{k=0:a''=a_0''+kr}^{\lfloor\frac{q}{r}\rfloor-1} e^{\frac{2i\pi}{q}a'(a_0''+kr)}\right|^2 =$$

$$\sum_{i'=0}^{r-1}\sum_{a':a'=nq/r} \frac{1}{q^2}\left|\sum_{k=0}^{\lfloor\frac{q}{r}\rfloor-1} e^{\frac{2i\pi}{q}a'kr}\right|^2 =$$

$$\sum_{n=0}^{r-1}\sum_{i'=0}^{r-1} \frac{1}{q^2}\left|\sum_{k}^{\lfloor\frac{q}{r}\rfloor-1} e^{\frac{2i\pi}{q}(nq/r)kr}\right|^2 =$$

$$\sum_{n=0}^{r-1}\sum_{i'=0}^{r-1} \frac{1}{q^2}\left|\sum_{k}^{\lfloor\frac{q}{r}\rfloor-1}\right|^2 = \sum_{n=0}^{r-1}\sum_{i'=0}^{r-1} \frac{1}{q^2}\lfloor\frac{q}{r}\rfloor^2 =$$

$$r \cdot r \cdot \frac{1}{q^2} \cdot \lfloor\frac{q}{r}\rfloor^2 \geq \frac{r^2}{q^2}(\frac{q}{r}-1)^2 = 1 - \frac{2r}{q} + \frac{r^2}{q^2} \geq 1 - \frac{2r}{q} \geq 1 - \frac{2N}{N^2} = 1 - \frac{2}{N}$$

We see that the output from the algorithm, with a high probability, will be some multiple of $\frac{q}{r}$. Using the chained fractions method, we know that we can determine the $r$ from this number with a quite good accuracy, see again [14]. As in the normal Shor algorithm, this must be done by some classical postprocessing. The postprocessing will be contained in our accepting function $a((x,N),y)$, which will then check that the $i$-th bit is 1. To feed the function with more multiples of $\frac{q}{r}$ at the same time, we may simply do a tensor product of any polynomial amount of our matrices $T$ and redefine the functions $c_I$ and $a$ in an analogous way. We know that there exists a polynomial $p(\text{length}(N))$ such that if we feed the function with $p(\text{length}(N))$ values, then

$$P(\text{we obtain correct } r \mid \text{given } p(\text{length}(N)) \text{ of multiples of } q/r) \geq \frac{3}{4}$$

That means that

$$P(\text{we obtain correct } r) =$$
$$P(\text{we obtain correct } r \mid \text{given } p(\text{length}(N)) \text{ of multiples of } q/r)$$
$$P(\text{given } p(\text{length}(N)) \text{ of multiples of } q/r) +$$
$$P(\text{we obtain correct } r \mid \text{given something else})P(\text{given something else})$$
$$\geq P(\text{we obtain correct } r \mid \text{given } p(\text{length}(N)) \text{ of multiples of } q/r) \cdot (1 - \frac{2}{N})^{p(\text{length}(N))}$$
$$+ P(\text{we obtain correct } r \mid \text{given something else}) \cdot (\frac{2}{N})^{p(\text{length}(N))} \geq$$
$$\geq \frac{3}{4} \cdot (1 - \frac{2}{N})^{p(\text{length}(N))}$$
$$\geq \frac{3}{4}(1 - p(\text{length}(N))\frac{2}{N})$$

which is greater than $\frac{2}{3}$ for $N > \frac{2}{9}P(\text{length}(N))$. There are only constantly many $N$ for which this inequality is not satisfied. $\square$

## 5.5 The class $M^2$

In this section, we will loosen Definition 5.5 even more and define yet another new complexity class, which will however turn out to be equal to another already a known class. This will close from above the hierarchy presented here.

**Definition 5.11** ($M^{t(n)}$). *A language $L$ is in $M^{t(n)}$ iff there is a series $T_1, T_2, \ldots$ (which does not have to be unitary) having property P2 and functions $c_I(x)$ and $c_A(x)$ computable in polynomial time, such that:*

*For $x \in L$, we have $1 \geq |\langle c_A(x)|T^{t(n)}|c_I(x)\rangle|^2 \geq \frac{2}{3}$.*

*For $x \notin L$, we have $0 \leq |\langle c_A(x)|T^{t(n)}|c_I(x)\rangle|^2 \leq \frac{1}{3}$.*

**Lemma 5.22.** *Let us have a language $L \in M^{2n}$ and a matrix $T$ and functions $c_I(x)$ and $c_A(x)$ solving it. Then there exists a matrix $S$ whose entries are computable in polytime and functions $c'_I(x)$ and $c'_A(x)$ computable in polytime such that $\langle c_A(x)|T^{2n}|c_I(x)\rangle = \langle c'_A(x)|S^2|c'_I(x)\rangle$.*

*Proof.* We extend the matrix $T$ adding one line to the top and one line to the left, filling them with zeros except the top left corner, where 1 will sit. Formally, we will have a new matrix $T'$ such that

$$\langle 0|T'|x\rangle \stackrel{\text{DEF}}{=} \langle x|T'|0\rangle \stackrel{\text{DEF}}{=} \delta_{x0}$$
$$\langle x|T'|y\rangle \stackrel{\text{DEF}}{=} \langle x-1|T|y-1\rangle$$

Let the size of the matrix $T'$ be $N$. The size of the new matrix $S$ will be $N(2n-1)$ Now we are ready to define the matrix $S$ as

$$\langle \vec{a}|S|\vec{b}\rangle \stackrel{\text{DEF}}{=} \langle a_1|T'|a_2\rangle\langle a_3|T'|a_4\rangle \ldots \langle a_{4n-3}|T'|a_{4n-2}\rangle$$
$$\langle a_{4n-1}|T'|b_1\rangle\langle b_2|T'|b_3\rangle \ldots \langle b_{4n-2}|T'|b_{4n-1}\rangle$$

and functions $c'_A(x) \stackrel{\text{DEF}}{=} |0, \ldots, 0, c_A(x)+1\rangle$ and $c'_I(x) \stackrel{\text{DEF}}{=} |c_I(x)+1, 0, \ldots, 0\rangle$ to get

$$\langle c'_A(x)|S^2|c'_I(x)\rangle = \langle 0,\ldots,0,c_A(x)+1|S^2|c_I(x)+1,0,\ldots,0\rangle$$
$$= \sum_{\vec{i}}\langle 0,\ldots,0,c_A(x)+1|S|\vec{i}\rangle\langle\vec{i}|S|c_I(x)+1,0,\ldots,0\rangle$$
$$= \sum_{\vec{i}}\langle 0|T'|0\rangle^{n-1}\langle c_A(x)+1|T'|i_1\rangle\langle i_2|T'|i_3\rangle\langle i_4|T'|i_5\rangle\ldots\langle i_{2n-2}|T'|i_{2n-1}\rangle$$
$$\langle i_1|T'|i_2\rangle\langle i_3|T'|i_4\rangle\ldots\langle i_{2n-3}|T'|i_{2n_2}\rangle\langle i_{4n-1}|T'|c_I(x)+1\rangle\langle 0|T'|0\rangle^{2n-1}$$
$$= \langle 0|T'|0\rangle^{2n-2}\sum_{\vec{i}}\langle c_A(x)+1|T'|i_1\rangle\langle i_1|T'|i_2\rangle\langle i_2|T'|i_3\rangle\ldots$$
$$\ldots\langle i_{2n-2}|T'|i_{2n-1}\rangle\langle i_{2n-1}|T'|c_I(x)+1\rangle$$
$$= \langle 0|T'|0\rangle^{2n-2}\langle c_A(x)+1|T'^{2n}|c_I(x)+1\rangle = \langle c_A(x)|T^{2n}|c_I(x)\rangle$$

□

**Corollary 5.23.** $M^p = M^2$ *for any polynomial $p$.*

*Proof.* For even values of the polynomial $p$, Lemma 5.22 can be used. For odd values the following trick can be used: Having a matrix $M$ that serves to show that a language $L$ is in $M^p$, we define for each integer $k$ a new matrix

$$S(k) \stackrel{\text{DEF}}{=} \begin{pmatrix} Me^{\frac{-\ln k}{k-1}} & 1 \\ 0 & Me^{\frac{-\ln k}{k-1}} \end{pmatrix}$$

Then it can be verified that for any integer $k \geq 1$ we have

$$S^k = \begin{pmatrix} \frac{M}{k}e^{\frac{-\ln k}{k-1}} & M^{k-1} \\ 0 & \frac{M}{k}e^{\frac{-\ln k}{k-1}} \end{pmatrix}$$

Therefore, using the matrix $S$ instead of $M$, we can add one to the power and thus arrive at an even number in the exponent. We can still find the desired entries in the top right corner. □

**Corollary 5.24.** $BQ_2^{poly} \subseteq M^2$.

*Proof.* Follows directly from definitions 5.5,5.11 and the previous corollary. □

**Lemma 5.25 (Amplification for the $M^2$ class).** *For each $L \in M^2$ and each polynomial $q(n)$, it holds that $L \in M^2$ with completeness $1 - e^{-q(n)}$ and soundness $e^{-q(n)}$.*

*Proof.* The proof method from Lemma 5.9 may be applied analogously here. □

**Lemma 5.26 (Multiple accepting configurations for $M^2$).** *Let us have a language $L$ such that there exists a matrix $T$ and functions $c_I(x)$ and $a(x,y)$ computable in polynomial time such that:*

*For $x \in L$, we have $1 \geq |\sum_{y:a(x,y)=1}\langle y|T^{t(n)}|c_I(x)\rangle|^2 \geq \frac{2}{3}$.*

*For $x \notin L$, we have $0 \leq |\sum_{y:a(x,y)=1}\langle y|T^{t(n)}|c_I(x)\rangle|^2 \leq \frac{1}{3}$.*

*Then $L \in M^2$.*

*Proof.* We need two functions $c'_I(x)$ and $c'_A(x)$ computable in polytime, a polynomial $q(n)$ and a matrix $M$ such that

For $x \in L$, we have $1 \geq \left|\langle c'_A(x)|M^{q(n)}|c'_I(x)\rangle\right|^2 \geq \frac{2}{3}$.

For $x \notin L$, we have $0 \leq \left|\langle c'_A(x)|M^{q(n)}|c'_I(x)\rangle\right|^2 \leq \frac{1}{3}$.

Let $T$, $c_I$ and $a(x,y)$ be the items as in Definition 5.6 that give the evidence for $L \in MQ_2^{p(n)}$. We will use the amplification in lemma 5.25 and assume we have

For $x \in L$: $\sum_{y:a(x,y)=1}\left|\langle y|T^{p(n)}|c_I(x)\rangle\right|^2 \geq \frac{8}{9}$.

For $x \notin L$: $\sum_{y:a(x,y)=1}\left|\langle y|T^{p(n)}|c_I(x)\rangle\right|^2 \leq \frac{1}{9}$.

To capture multiple accepting configurations into the notion of the class $M^2$, we will use the following trick: We will obtain the sum

$$\sum_{y:a(x,y)=1}\left|\langle y|T^{p(n)}|c_I(x)\rangle\right|^2$$

in a matrix element of $M^{p+1}$. The matrix $M$ will have three indices, $x, x'$ and $y$. The third index will be used to count the number of elements. In the expression for $M^{p+1}$, we will want to have $p$ times pairs of elements of $T$ and one time a pair of indicators whether a corresponding pair of elements of $T$ is accepting or not. Therefore, in the space of the third index, the matrix will have nonzero elements only on the main diagonal shifted up by 1, which will cause that in the expression of $M^{p+1}$, each element of $M$ will have the index of the following one plus 1. Thus, if we start with 1, the first element will have index $p + 1$. This element will be 1 if corresponding element in $T$ has the function $a$ equal to 1, or zero otherwise.

Formally, we define a matrix $M$ and functions $c'_I(x)$, $c'_A(x)$ as

$$\langle x_1, x'_1, y_1|M|x_2, x'_2, y_2\rangle \stackrel{\text{DEF}}{=} \begin{cases} \delta_{y_1,y_2+1}\langle x_1|T|x_2\rangle\langle x'_1|T^*|x'_2\rangle & \text{if } y_2 \neq p \\ \delta_{y_1,y_2+1}\delta_{x_2,x'_2} & \text{if } a(x,x_2)=1 \\ & \text{and } y_2=p \\ 0 & \text{otherwise} \end{cases}$$

$$c'_I(x) \stackrel{\text{DEF}}{=} |c_I(x), c_I(x), 1\rangle$$
$$c'_A(x) \stackrel{\text{DEF}}{=} |0, 0, p+1\rangle$$

Then we have

$$\langle c_A(x)|M^{p+1}|c_I(x)\rangle = \langle 0,0,p+1|M^{p+1}|c_I(x),c_I(x),1\rangle =$$
$$\sum_{\vec{i},\vec{j},\vec{k}} \langle 0,0,p+1|M|i_1,j_1,k_1\rangle\langle i_1,j_1,k_1|M|i_2,j_2,k_2\rangle\ldots\langle i_p,j_p,k_p|M|c_I(x),c_I(x),1\rangle$$
$$= \sum_{\vec{i},\vec{j}} \sum_{k_1,k_2\ldots k_{p-1}} \langle 0,0,p+1|M|i_1,j_1,k_1\rangle\langle i_1,j_1,k_1|M|i_2,j_2,k_2\rangle\ldots$$
$$\ldots \langle i_{p-1},j_{p-1},k_{p-1}|M|i_p,j_p,2\rangle\langle i_p,j_p,2|M|c_I(x),c_I(x),1\rangle$$
$$=\ldots=$$
$$\sum_{\vec{i},\vec{j}} \langle 0,0,p+1|M|i_1,j_1,p\rangle\langle i_1,j_2,p|M|i_2,j_2,p-1\rangle\ldots$$
$$\ldots \langle i_{p-1},j_{p-1},2|M|i_p,j_p,2\rangle\langle i_p,j_p,2|M|c_I(x),c_I(x),1\rangle$$
$$=\ldots=$$
$$\sum_{\vec{i}:a(x,i_1)=1} 1\cdot\langle i_1|T|i_2\rangle\ldots\langle i_{p-1}|T|i_p\rangle\langle i_p|T|c_I(x)\rangle$$
$$\sum_{\vec{j}:j_1=i_1} 1\cdot\langle j_1|T^*|j_2\rangle\ldots\langle j_{p-1}|T^*|j_p\rangle\langle j_p|T^*|c_I(x)\rangle$$
$$\sum_{i_1:a(x,i_1)=1} \langle i_1|T^p|c_I(x)\rangle\langle i_1|T^{*p}|c_I(x)\rangle = \sum_{i_1:a(x,i_1)=1} |\langle i_1|T^p|c_I(x)\rangle|^2$$

We obtained

For $x \in L$:

$$\left|\langle c_A(x)|M^{p+1}|c_I(x)\rangle\right|^2 = \left|\sum_{i_1:a(x,i_1)=1}\left|\langle i_1|T^p|c_I(x)\rangle\right|^2\right|^2 \geq \left(\frac{8}{9}\right)^2 \geq \frac{2}{3}$$

For $x \notin L$:

$$\left|\langle c_A(x)|M^{p+1}|c_I(x)\rangle\right|^2 = \left|\sum_{i_1:a(x,i_1)=1}\left|\langle i_1|T^p|c_I(x)\rangle\right|^2\right|^2 \leq \left(\frac{1}{9}\right)^2 \leq \frac{1}{3}$$

Due to the assumption, we also have that $\sum_{i_1:a(x,i_1)=1}\left|\langle i_1|T^p|c_I(x)\rangle\right|^2 \leq 1$. So the matrix $M$ shows that $L \in M^{p(n)+1}$. Using lemma 5.22, it follows that $L \in M^2$. □

**Corollary 5.27.** $MQ_2^{poly} \subseteq M^2$

*Proof.* Follows the definition of 5.6, 5.11 and the previous lemma. □

**Lemma 5.28 (Real numbers in $M^2$).** *Let us have a language $L \in M^2$. Then there exist a matrix $M$ documenting that $L \in MQ_2^{t(n)}$ such that there are only real numbers as the entries of $M$.*

*Proof.* Let $T$, $c_I(x)$ and $c_A(x)$ be the witnesses of $L \in M^2$. We define

$$M \stackrel{\text{DEF}}{=} \begin{pmatrix} Re & -Im \\ Im & Re \end{pmatrix}$$
$$|c_I'(x)\rangle \stackrel{\text{DEF}}{=} |c_I(x),0\rangle$$
$$a'(x,(y,y')) \stackrel{\text{DEF}}{=} 1 \iff y = c_A(x)$$

where $Re \overset{\text{DEF}}{=} Re(T)$ and $Im \overset{\text{DEF}}{=} Im(T)$. Then we get

$$M^2 = \begin{pmatrix} Re^2 - Im^2 & -ImRe - ReIm \\ ImRe + ReIm & Re^2 - Im^2 \end{pmatrix}$$

For each $x, y$, we may write

$$\langle x|T^2|y\rangle = \langle x|\,(Re + iIm))^2\,|y\rangle = \langle x|(Re^2 - Im^2)|y\rangle + i\langle x|(ReIm + ImRe)|y\rangle$$

and so we have

$$\left|\langle x|T^2|y\rangle\right|^2 = \left|\langle x|(Re^2 - Im^2)|y\rangle\right|^2 + \left|\langle x|(ReIm + ImRe)|y\rangle\right|^2$$

This means that the norm of pairs of entries in the corresponding positions in the submatrices in the first column is equal to the norm of the corresponding entry in the original matrix $T$. We get

$$\sum_{(y,y'):a(x,(y,y'))=1} \left|\langle y, y'|M^2|c_I'(x)\rangle\right|^2 =$$

$$\left|\langle c_A(x), 0|M^2|c_I(x), 0\rangle\right|^2 + \left|\langle c_A(x), 1|M^2|c_I(x), 0\rangle\right|^2 = \left|\langle c_A(x)|T^2|c_I(x)\rangle\right|^2$$

The proof is finished using Lemma 5.26, since the construction there does not introduce any new complex numbers. $\square$

**Lemma 5.29 (Precision required in class $M^2$).** *Let us have a language $L \in M^2$. Then there exist a matrix $M$ documenting that $L \in MQ_2^{t(n)}$ such that for some polynomial $q(n)$, all elements in the matrix $M$ are exact multiples of $\frac{1}{2^{q(n)}}$.*

*Proof.* Let $T$ be a witness of $L \in M^2$ and let the size of the matrix $T$ be $2^{p(n)}$ for $p$ being a polynomial. We know from lemma 5.28 that we may assume there are only real entries in the matrix $T$. We define $M$ as

$$\langle i|M|j\rangle \overset{\text{DEF}}{=} \text{sign}(\langle i|T|j\rangle) \cdot \frac{1}{2^{q(n)}} \left\lfloor 2^{q(n)}|\langle i|T|j\rangle| \right\rfloor$$

Then the biggest error we may get when computing the result over an input is

$$\left|\langle i|T^2|j\rangle - \langle i|M^2|j\rangle\right| \leq 2^{p(n)} \cdot \frac{2}{2^{q(n)}} = 2^{1+p(n)-q(n)}$$

because when computing an entry of $T^2$, we are summing over one index with $2^{p(n)}$ values and the items in the sum are products of 2 items of precision $\frac{1}{2^{q(n)}}$. Setting $q(n) \geq p(n) + 4$, we get that the error is at most $\frac{1}{8}$. Using the amplification lemma 5.25, we can get back the numbers $\frac{1}{3}$ and $\frac{2}{3}$. Furthermore, since $|\langle i|M|j\rangle| \leq |\langle i|T|j\rangle|$, we have that

$$\left|\langle i|M^2|j\rangle\right| \leq \left|\langle i|T^2|j\rangle\right| \leq 1$$

$\square$

In the following lemma, we provide a characterization of the class $M^2$, which we will use later on.

**Lemma 5.30 ($M^2$ characterization).** $L \in M^2 \iff$ *there exists a function $G(x,y) : (\{0,1\}^n, \{0,1\}^n) \to \{-1,1\}$ computable in polynomial time and a polynomial $p(n)$ such that for each $x$:*

*if $x \in L$ then $|\sum_y G(x,y)|^2 \geq 2^{p(n)} \frac{2}{3}$,*

if $x \notin L$ then $|\sum_y G(x,y)|^2 \leq 2^{p(n)}\frac{1}{3}$,

where $n = length(x)$. Furthermore, it holds that $\frac{|\sum_y G(x,y)|^2}{2^{p(n)}} \leq 1$.

*Proof.* Using the setting and notation from the definition of $M^2$, let $n$ be the length of $c_I(x)$. We define the function $F(x,k)$ as

$$F(x,k) \equiv \langle c_A(x)|T|k\rangle\langle k|T|c_I(x)\rangle$$

and write

$$|\langle c_A(x)|T^2|c_I(x)\rangle|^2 = |\sum_{k=1}^{2^n} \langle c_A(x)|T|k\rangle\langle k|T|c_I(x)\rangle|^2 = |\sum_{k=1}^{2^n} F(x,k)|^2 \leq 1$$

The inequality follows from the definition of the class $M^2$ (Definition 5.11). The range of the functions $F(x,k)$ is the interval $[-1,1]$. Using lemma 5.29, we know that we may assume there exists a polynomial $q(n)$ such that the functions $F(x,k)$ output multiples of $\frac{1}{2^{q(n)}}$. Now, we may define function $G(x,(y,z)) : (\{0,1\}^n, (\{0,1\}^n, \{0,1\}^{q(n)})) \to \{0,1\}$ as

$$G(x,(y,z)) \stackrel{\text{DEF}}{=} \begin{cases} 1 & \text{if } F(x,y) \geq 0 \ \& \ F(x,y) \geq \frac{z}{2^{q(n)}} \\ -1 & \text{if } F(x,y) < 0 \ \& \ F(x,y) \leq \frac{-z}{2^{q(n)}} \\ & \text{otherwise decide with probability } \frac{1}{2} \text{ between } -1 \text{ and } 1 \end{cases}$$

Then we have that

$$\sum_z G(x,(y,z)) = F(x,y)2^{q(n)}$$

and thus

$$\left|\sum_{k=1}^{2^n} \sum_z G(x,(k,z))\right|^2 = 2^{2q(n)}\left|\sum_{k=1}^{2^n} F(x,k)\right|^2$$

Setting $p(n) = 2q(n)$ proves the lemma. □

## 5.6 GapP functions

In this section, we will present a definition of the so called *GapP functions* defined in [13], and the class $AWPP$ which is defined using these functions. Then we will prove $M^2 = AWPP$.

**Definition 5.12 (Gap function).** *A function $f$ is in the class Gap iff there exists a non-deterministic Turing machine $M$ running in polynomial time such that $f(x)$ is the number of accepting paths of $M$ over $x$ minus the number of rejecting paths of $M$ over $x$.*

In [13], the following lemma is shown:

**Lemma 5.31 (Gap functions properties).** *Let $f(x) \in Gap$ and $q$ a polynomial. Then the following are functions in Gap:*

1. $-f(x)$,,

2. $\sum_{|y|<q(|x|)} f(x,y)$,

3. $\prod_{y<q(|x|)} f(x,y)$.

We will use also the following, very handy characterization, proved in [8].

**Lemma 5.32 (AWPP characterization).** *A language $L$ is in $AWPP$ iff there exists a polynomial $p$ and a function $g \in Gap$ such that:*

*For $x \in L$, we have $\frac{2}{3} \leq g(x)/2^p \leq 1$.*

*For $x \notin L$, we have $0 \leq g(x)/2^p \leq \frac{1}{3}$.*

The authors also showed that the constants $\frac{1}{3}$ and $\frac{2}{3}$ may be amplified to $2^{-n}$ and $1 - 2^{-n}$ respectively.

**Lemma 5.33 (Characterization of $BPP$ using $GapP$).** *Let us have a language $L \in AWPP$ and $g(x)$ and $p$ as in Lemma 5.32, showing that $L \in AWPP$, and a nondeterministic Turing machine $M(x, y)$ such that its gap is exactly $g(x)$ and the length of $y$ is exactly $p$. Then $L \in BPP$.*
*Conversely, for each $L \in BPP$, we can find such $g(x)$ and $M(x, y)$.*

*Proof.* First, let us have a language $L$, $g(x)$ and $M(x, y)$ as above and let us denote $M(x, y) = 1$ if $M$ over input $x$ with random bits $y$ accepts and $M(x, y) = 0$ otherwise. Then

$$g(x) = \sum_{y:M(x,y)=1} 1 - \sum_{y:M(x,y)=0} 1$$

and according to the assumption about the length of $y$:

$$2^p = \sum_{y:M(x,y)=1} 1 + \sum_{y:M(x,y)=0} 1.$$

Thus, solving these two equations, we get that:

If $g(x)/2^p \geq \frac{2}{3}$ then $P(M \text{ accepts } x) = \frac{\sum_{M(x,y)=1} 1}{2^p} \geq \frac{1}{2} + \frac{1}{3}$ and

if $g(x)/2^p \leq \frac{1}{3}$ then $P(M \text{ accepts } x) = \frac{\sum_{M(x,y)=1} 1}{2^p} \leq \frac{1}{2} + \frac{1}{6}$.

Using Lemma 5.3, we get that $L \in BPP$.

On the other side, if we have $L \in BPP$ and a machine $M(x, y)$ with $p$ random bits accepting it, then

For $x \in L$: $1 \geq \frac{g(x)}{2^p} \geq \frac{2}{3} - \frac{1}{3} = \frac{1}{3}$,

for $x \notin L$: $\frac{g(x)}{2^p} \leq \frac{1}{3} - \frac{2}{3} = -\frac{1}{3}$.

We will therefore add another two bits $b_1, b_2$ to the machine $M$ and define:

$$M(x, y, b_1, b_2) \stackrel{\text{DEF}}{=} \begin{cases} 1 & \iff M(x, y) = 1 \lor b_1 \neq b_2 \\ 0 & \text{otherwise} \end{cases}$$

Then $P(M \text{ accepts } x) = 1 - P_y(M(x, y) = 0)\frac{1}{2}$ and thus

For $x \in L$: $1 \geq \frac{g(x)}{2^p} \geq \left(1 - \frac{1}{3}\frac{1}{2}\right) - \left(\frac{1}{3}\frac{1}{2}\right) = \frac{2}{3}$,

For $x \notin L$: $0 \leq \frac{g(x)}{2^p} \leq \left(1 - \frac{2}{3}\frac{1}{2}\right) - \left(\frac{2}{3}\frac{1}{2}\right) = \frac{1}{3}$,

□

**Theorem 5.34.** $AWPP \supseteq M^2$

*Proof.* Using Lemma 5.30, we know that for each language $L \in M^2$ there exists a function $G(x, y)$ computable in polytime and a polynomial $p(n)$ such that for each $x$:

if $x \in L$ then $|\sum_y G(x, y)|^2 \geq 2^{p(n)}\frac{2}{3}$,

if $x \notin L$ then $|\sum_y G(x, y)|^2 \leq 2^{p(n)}\frac{1}{3}$,

where $n = length(x)$. Since the range of $G$ is only $\{1, -1\}$, we have that $G \in GapP$. Using Lemma 5.31, we have that $|\sum_y G(x,y)|^2 \in Gap$ as well. It remains to prove $\frac{|\sum_y G(x,y)|^2}{2^{p(n)}} \leq 1$, which is also thanks to Lemma 5.30. □

**Theorem 5.35.** $AWPP \subseteq M^2$.

*Proof.* Let us have a language $L \in AWPP$, a function $g(x) \in GapP$, and a polynomial $p$ proving that $L \in AWPP$ as in Lemma 5.32. We assume that $g$ and $p$ are amplified to

For $x \in L$, we have $\frac{8}{9} \leq g(x)/2^p \leq 1$.

For $x \notin L$, we have $0 \leq g(x)/2^p \leq \frac{1}{9}$.

Let us now have a nondeterministic Turing machine $M(x, y)$ for which the function $g(x)$ computes the gap. We then define a matrix $M$ as

$$\langle x, y|M|x', y'\rangle \stackrel{\text{DEF}}{=} 2^{-p/2} \delta_{x,x'} (-1)^{\sum_i y_i, y'_i} (-1)^{M(x,y)}$$

Such a matrix is furthermore orthogonal, since its proportional to the Hadamard matrix working on the second register with some rows multiplied by -1, which does not change the unitarity. But here we will not need this fact. We get

$$\langle x, 0|M^2|x, 0\rangle = \sum_{i,j} \langle x, 0|M|i, j\rangle \langle i, j|M|x, 0\rangle =$$

$$\sum_j \langle x, 0|M|x, j\rangle \langle x, j|M|x, 0\rangle =$$

$$\sum_j 2^{-p} (-1)^{\sum_i 0_i, j_i} (-1)^{M(x,0)} (-1)^{\sum_i j_i, 0_i} (-1)^{M(x,j)}$$

$$= 2^{-p} (-1)^{M(x,0)} \sum_j (-1)^{M(x,j)} = \frac{g(x)}{2^p}$$

And thus

For $x \in L$, we have $1 \geq |\langle x, 0|M^2|x, 0\rangle|^2 = \left(\frac{g(x)}{2^p}\right)^2 \geq \left(\frac{8}{9}\right)^2 \geq \frac{2}{3}$.

For $x \notin L$, we have $0 \leq |\langle x, 0|M^2|x, 0\rangle|^2 = \left(\frac{g(x)}{2^p}\right)^2 \leq \left(\frac{1}{9}\right)^2 \leq \frac{1}{3}$.

This finishes the proof. □

The following theorem is a corollary of Observation 5.10 saying that $BPP \subseteq MQ^1$ and Theorem 5.16 saying $MQ^1 \subseteq BQ^2$. Nevertheless, a direct proof of $BPP \subseteq BQ^2$ is much easier and elegant:

**Theorem 5.36.** $BPP \subseteq BQ^2$.

*Proof.* We proceed exactly as in the proof of Theorem 5.35. Using Lemma 5.33, we know that there is a machine $M(x, y)$ with exactly $p$ random bits. Thus the matrix $M$ defined as

$$\langle x, y|M|x', y'\rangle \stackrel{\text{DEF}}{=} \sqrt{\frac{1}{2^p}} \delta_{x,x'} (-1)^{\sum_i y_i, y'_i} (-1)^{M(x,y)}$$

is properly normalized, since there are exactly $2^p$ different $y$'s and so each row and column has a unit norm. The matrix is furthermore unitary, since $(-1)^{\sum_i y_i, y'_i}$ is an entry of a Hadamard matrix and the element $(-1)^{M(x,y)}$ multiplies some columns by $-1$. Multiplying some columns by $-1$ obviously can not change orthogonality of columns, but it can neither change the orthogonality of rows, since always both the corresponding entries in the rows have their sign changed and thus in the inner product the sings will cancel. We then conclude that the matrix $M$ is unitary. □

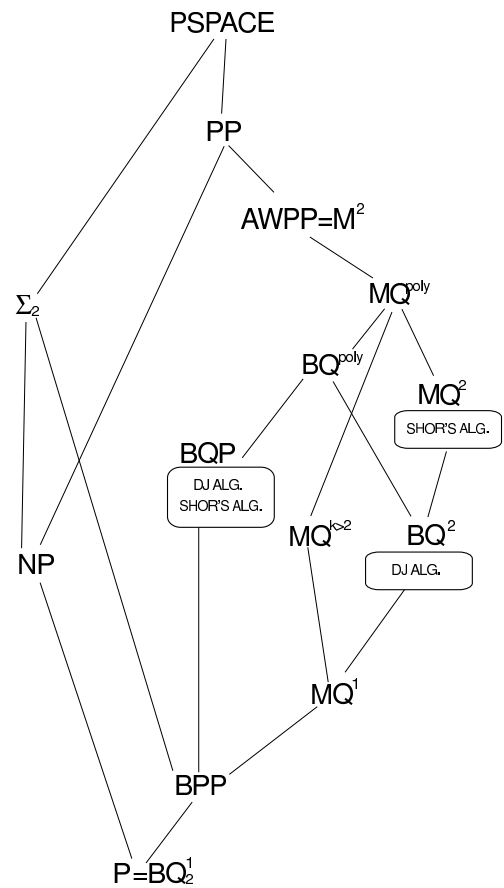

Figure 5.5: Hierarchy of classes including some new ones. For each pair connected by a line, the class that stays upper contains the lower one. $MQ^{k>2}$ denotes any class $MQ^k$ for a constant $k > 2$.

## 5.7 Hierarchy with the new classes

The hierarchy is visualised on figure 5.7.

# Conclusions and remarks

We summarized the concept of Quantum Computing, both from the view of quantum Turing machines and the view of Quantum circuits. We also showed the relation between the two models and touched the phenomenon of entanglement. Then we reviewed some properties of the class $BQP$, a class of problems that are efficiently solvable by quantum Turing machines and quantum circuits.

In Chapter 5, we define a new bunch of classes inspired by the quantum world and the class $BQP$. Their definitions are natural generalizations or modifications of the definition of $BQP$. One part was obtained by changing the condition that a circuit be of a polynomial size into the condition that each element of a matrix representing a circuit be a polynomial time computable function and we may have more identical matrices in the circuit. So we arrived at a bunch of classes $BQ^{t(n)}$ for $t(n)$ being some function that tells us how many identical matrices in the circuit we have for an input of length $n$. This definition is very easy to work with. Generalizing this definition to the possibility of more accepting configurations, we arrived at the family $MQ^{t(n)}$. The relation between the classes is that for each $i$: $BQ^i \subseteq MQ^i$. All these classes form a huge hierarchy and are all contained in the class $AWPP$. However, none of the inclusions is known to collapse, nor to be proper. Showing a proper inclusion would immediately lead to $P \neq PSPACE$, which is considered to be a very hard question.

From above, it followed that $BQ^{poly(n)} \supseteq BQP$. Nevertheless, we showed that the two main quantum algorithms, Deutsch-Jozsza and Shor algorithm belong to the class $BQ^2$ and $MQ^2$ respectively, which raises the question whether the $poly(n)$ in $BQ^{poly(n)}$ can be bounded for example to be a linear function or a constant, or at least to put $BQP \subseteq MQ^{t(n)}$ with $t(n)$ being something smaller than a general polynomial.

To come back to the classical world, we loosened the definitions of the classes $MQ^{t(n)}$ and $BQ^{t(n)}$ even more and dropped the unitarity condition of the matrices to arrive at the class $M^2$ which is shown to be equal to the class $AWPP$ defined by [10]. This class was invented as an artificial class meant to be used just for proving lowness theorem of $BQP$. Here we showed that its definition was surprisingly natural.

The advantage of all these new classes is that they sit in the hierarchy between $BQP$, $BPP$ and $AWPP$ and are $\Sigma_2$ definable with easy definitions resembling the definition of $BPP$ and $BQP$, and so fill the gap between the wellknown classes $BQP$ and $AWPP$ in a hierarchy where proving many things is considered to be very hard, such as whether $AWPP = BQP$. The classes suggested here therefore leave place for future proofs which might lead to showing some properties of $BQP$ itself. They may also help to understand better how the unitarity of the nature influences computation in general.